\documentclass[journal=jpclcd,manuscript=letter,layout=twocolumn]{achemso}

\usepackage[version=3]{mhchem} 
\usepackage{lmodern}
\usepackage{color}
\usepackage[dvipsnames,svgnames,table]{xcolor}
\usepackage[colorlinks=true,linkcolor=blue,urlcolor=blue,citecolor=blue]{hyperref}
\usepackage{graphicx}
\usepackage{dcolumn}
\usepackage{array}
\usepackage{bm}
\usepackage{subfigure}
\usepackage{amssymb}
\usepackage{multirow}
\usepackage{tabularx}
\usepackage{amsmath}
\usepackage{makecell}

\newcommand{\deri}[2]{\frac{\partial #1}{\partial #2}}
\newcommand{\derii}[2]{\frac{\partial^2 #1}{\partial #2^2}}
\renewcommand{\vec}[1]{\boldsymbol{#1}}
\newcommand{\mat}[1]{\mathbf{#1}}
\DeclareMathSymbol{\shortminus}{\mathbin}{AMSa}{"39}
\newcommand{\erfc}{\text{erfc}}
\newcommand{\boltz}{k_{\text{B}}}
\newcommand{\bohr}{a_{\text{B}}}
\newcommand{\order}[1]{\mathcal{O}\left(#1\right)}
\DeclareMathOperator*{\argmax}{arg\,max}
\DeclareMathOperator*{\argmin}{arg\,min}

\iftrue
    \newcommand{\todo}[1]{{\color{red}[#1]}} 
    \newcommand{\question}[1]{{\color{blue}[#1]}}
\else
    \newcommand{\todo}[1]{}               
    \newcommand{\question}[1]{}
\fi

\author{Pontus Svensson}
\email{p.svensson@hzdr.de}
\affiliation{Center for Advanced Systems Understanding (CASUS), D-02826 G\"orlitz, Germany}
\alsoaffiliation{Helmholtz-Zentrum Dresden-Rossendorf (HZDR), D-01328 Dresden, Germany}

\author{Fotios Kalkavouras}
\affiliation{Space and Plasma Physics, Royal Institute of Technology (KTH), Stockholm, SE-100 44, Sweden}

\author{Uwe Hernandez Acosta}
\affiliation{Center for Advanced Systems Understanding (CASUS), D-02826 G\"orlitz, Germany}
\alsoaffiliation{Helmholtz-Zentrum Dresden-Rossendorf (HZDR), D-01328 Dresden, Germany}

\author{Zhandos~A.~Moldabekov}
\affiliation{Center for Advanced Systems Understanding (CASUS), D-02826 G\"orlitz, Germany}
\alsoaffiliation{Helmholtz-Zentrum Dresden-Rossendorf (HZDR), D-01328 Dresden, Germany}

\author{Panagiotis Tolias}
\affiliation{Space and Plasma Physics, Royal Institute of Technology (KTH), Stockholm, SE-100 44, Sweden}

\author{Jan Vorberger}
\affiliation{Helmholtz-Zentrum Dresden-Rossendorf (HZDR), D-01328 Dresden, Germany}

\author{Tobias Dornheim}
\affiliation{Center for Advanced Systems Understanding (CASUS), D-02826 G\"orlitz, Germany}
\alsoaffiliation{Helmholtz-Zentrum Dresden-Rossendorf (HZDR), D-01328 Dresden, Germany}

\title{Accelerated free energy estimation in \emph{ab initio} path integral Monte Carlo simulations}

\keywords{Free energy, PIMC, Warm dense matter, Uniform electron gas}

\begin{document}

\begin{tocentry}
    \includegraphics[width=\linewidth]{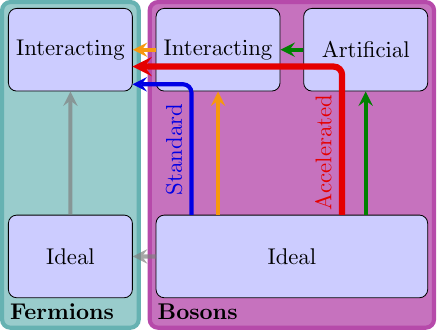}
\end{tocentry}

\begin{abstract}
    We present a methodology for accelerating the estimation of the free energy from path integral Monte Carlo simulations by considering an intermediate artificial reference system where interactions are inexpensive to evaluate numerically. Using the spherically averaged Ewald interaction as this intermediate reference system for the uniform electron gas, the interaction contribution for the free energy was evaluated up to 18 times faster than the Ewald-only method. Furthermore, a $\xi$-extrapolation technique was tested and applied to alleviate the fermion sign problem and to resolve the sign for large particle numbers. Combining these two techniques enabled the evaluation of the free energy for a system of 1000 electrons, where both finite-size and statistical errors are below chemical accuracy. The general procedure can be applied to systems relevant for planetary and inertial confinement fusion modeling with low to moderate levels of quantum degeneracy.
\end{abstract}

The description of thermal systems of interacting fermions is a cornerstone of our understanding for a wide range of quantum systems, including ultra-cold atoms~\cite{giorgini2008theory,ceperley1992path}, quantum dots~\cite{reimann2002electronic,dornheim2022abnormal}, and dense plasmas~\cite{lorenzen2014progress}. In particular, dense quantum plasmas are abundant in astrophysics, where they are found in gas giants~\cite{guillot1999interiors,fortney2010interior,helled2020understanding} such as Jupiter~\cite{french2012ab}, Saturn~\cite{preising2023material} and some classes of exoplanets~\cite{guillot1999interiors}, and stars~\cite{kippenhahn2012stellar} most notable in later stages of stellar evolution in the form of red giants~\cite{fortov2009extreme}, white dwarfs~\cite{chandrasekhar1931density,chabrier2000cooling} and the atmospheres of neutron stars~\cite{gudmundsson1983structure,haensel2007neutron}. However, high-density plasmas are also central in human-made applications such as inertial confinement fusion (ICF)~\cite{nuckolls1972laser,betti2016inertial,hurricane2023physics} and the synthesis of novel materials~\cite{miao2020chemistry}. In recent groundbreaking experiments, ICF implosions have exceeded the Lawson criteria and achieved capsule gain~\cite{abu2022lawson}, a key step towards achieving energy production through the ICF concept.

A formidable regime of dense plasmas to model theoretically is warm dense matter (WDM), which is characterised by a complex interplay between interactions, quantum degeneracy, and thermal excitations~\cite{lorenzen2014progress,bonitz2020ab,vorberger2025roadmapwarmdensematter}. All the previously mentioned effects must be taken into account as both $r_s$ -- the ratio between the Wigner-Seitz radius and the Bohr radius -- and $\theta$ -- the ratio of the thermal excitation energy and the electronic Fermi energy -- are of order unity, which characterises the strength of interactions and quantum degeneracy, respectively. Therefore, there remain uncertainties in the fundamental properties of WDM, such as the equation of state (EOS) and transport properties, which limit predictive modeling of, for example, the Jovian interior~\cite{wahl2017comparing,helled2020understanding} and ICF implosions~\cite{hu2015impact}.

The most widely used description for WDM systems is a hybrid method (DFT-MD)~\cite{bonitz2024principles}, where electrons are described using density functional theory (DFT)~\cite{hohenberg1964inhomogeneous,mermin1965thermal}, while ions are treated by molecular dynamics (MD)~\cite{rapaport2004art}. Formally, DFT is exact given the correct exchange-correlation functional ~\cite{hohenberg1964inhomogeneous}, but this functional remains unknown and practical calculations resort to approximate descriptions, often based on the properties of the uniform electron gas (UEG)~\cite{perdew1981self,karasiev2014accurate,dornheim2018uniform}. Path integral Monte Carlo (PIMC)~\cite{ceperley1995path,vorberger2025roadmapwarmdensematter,Bohme_PRL_2022} provides a suitable benchmark at finite temperature, since it is exact within the statistical error. However, for fermionic systems, PIMC is limited by the fermion sign problem (FSP) in the number of particles and the level of quantum degeneracy it can model~\cite{dornheim2019fermion}. The FSP arises because all fermionic observables are ratios where the denominator is the average sign $S$, which decreases exponentially with particle number and the inverse temperature~\cite{dornheim2019fermion,troyer}. This vanishing sign causes computations of large or cold systems to be dominated by statistical errors~\cite{hatano1994data}. 

To address the FSP, \citeauthor{xiong2022thermodynamic} have suggested a $\xi$-extrapolation method~\cite{xiong2022thermodynamic} where an additional $\xi$ parameter that smoothly interpolates from the bosonic ($\xi = 1$) to the fermionic ($\xi = -1$) limit was introduced. By introducing an empirical model for the $\xi$-dependence, calculations can be carried out in the FSP-free parameter regime and extrapolated to the fermionic results, circumventing the exponential computational cost with respect to the particle number~\cite{dornheim2023fermionic,Dornheim_JPCL_2024}. This extrapolation method has been successfully applied to moderately degenerate systems ($\theta \geq 1.0$) for the computation of energy~\cite{xiong2023thermodynamics,dornheim2023fermionic,morresi2025normal}, static structure~\cite{dornheim2023fermionic,dornheim2024ab,dornheim2024ab_b,dornheim2025unraveling}, imaginary time correlation function~\cite{dornheim2023fermionic,dornheim2025unraveling}, density response~\cite{dornheim2024ab,dornheim2024ab_c}, and the average sign itself~\cite{dornheim2025fermionic}.

The (Helmholtz) free energy is central for our understanding of thermal systems, for example it is directly related to the exchange-correlation functional in DFT~\cite{mermin1965thermal,gupta1982density,smith2018warming} where finite-temperature corrections are key at intermediate temperatures~\cite{sjostrom2014gradient,karasiev2016importance,ramakrishna2020influence,bonitz2024principles}, but the free energy is also commonly used to investigate the stability of different phases~\cite{alfe1999melting,wilson2011solubility,wu2021high}. As the free energy is a thermodynamic potential, a free energy parametrisation automatically yields a self-consistent EOS where all thermodynamic properties are obtained through differentiation. So far, first-principles tabulations of the EOS have focused on energy and pressure~\cite{militzer2021first}, but semi-empirical constructions commonly model the free energy~\cite{more1988new,lyon1995sesame}. By accelerating first-principle computations of the free energy, we are moving closer to reliable and internally consistent equation of state tables in the WDM regime. 

In this letter, we present computations for the free energy of the spin-unpolarised UEG from PIMC with unprecedentedly large system sizes and low statistical errors. Using a combination of robust extrapolation techniques and the introduction of an intermediate reference system where interactions are computationally cheap, we are able to model $N = 1000$ electrons. The efficiency of this scheme allows us to evaluate the free energy to well within chemical accuracy (i.e., $1\,\text{kcal/mol} \approx 1.6\,\text{mHa}$~\cite{pople1999nobel}). In the main text, we focus on the condition $r_s = 3.23$ and $\theta = 1.0$ characteristic of the electronic conditions possible to achieve in hydrogen jet experiments~\cite{Zastrau2021,Fletcher_Frontiers_2022,Hamann_PRR_2023}, but the methodology is general and applicable for either bosons and not too degenerate Fermi systems. The complete analysis for the UEG at $r_s = 10$ is given in the supporting information.

The partition function or the free energy is not a thermodynamical average \textit{per se}, but relates to a volume in phase space~\cite{frenkel2002understanding}. Therefore, the free energy is not readily available from an MC or MD simulation, and the thermodynamic integration (TI)~\cite{zwanzig1954high,frenkel2002understanding} method or the adiabatic connection (AC) formula~\cite{pribram2016thermal} has traditionally been used for its computation. Both methods require multiple computations, e.g., with an interaction that can be smoothly turned from that of a reference system -- commonly the ideal system -- to the target system. Moreover, the application of the AC method to inhomogeneous systems, such as the electronic problem in the external potential of a fixed ion configuration, poses an additional obstacle. Recently, \citeauthor{dornheim2025direct} introduced the extended ensemble technique in which the free energy differences between systems 1 and 2 can be directly computed~\cite{dornheim2025direct}. The extended partition function in question is
\begin{equation}
    Z_{\text{ext}} = c Z_1 + Z_2,
    \label{eq:extended_ensemble}
\end{equation}
where $Z_i$ is the partition function of system $i$, and $c$ is an arbitrary coefficient that is chosen to optimise the ergodicity~\cite{dornheim2025direct}. In the extended ensemble, the difference in free energy per particle $f_i$ between the two systems is directly related to the thermal averages in the extended ensemble $\langle \cdot \rangle_{\text{ext}}$ via
\begin{equation}
    f_1 - f_2 = -\frac{\boltz T}{N} \log\left( \frac{c^{-1} \langle \Hat{\delta_1} \rangle_{\text{ext}}}{\langle \Hat{\delta_2} \rangle_{\text{ext}}} \right),
    \label{eq:extended_ensemble_free_energy}
\end{equation}
where $\Hat{\delta}_i$ is one in system $i$ and zero otherwise, $\boltz T$ is the temperature in energy units, and $N$ is the number of particles.

The Hamiltonian $\Hat{H}_{\eta} = \Hat{K} + \eta \Hat{V}$ where $\Hat{K}$ is the kinetic energy operator and $\Hat{V}$ is the Ewald summation~\cite{ewald1921berechnung,dornheim2018uniform}, interpolates between the ideal ($\eta = 0$) and interacting systems ($\eta = 1$). By considering $\eta = 0$ and $\eta = 1$ for the two systems in Equation \eqref{eq:extended_ensemble_free_energy} along with exact results for noninteracting systems~\cite{zhou2018canonical,barghathi2020theory}, the free energy for bosons can be computed~\cite{dornheim2025direct,dornheim2025eta} in what we will refer to as the $\eta$-ensemble. However, for a large number of particles, it was found practically difficult to ergodically explore the entire extended ensemble due to the presence of configurations that are strongly suppressed for interacting systems in the ideal case~\cite{dornheim2025eta}. Therefore, multiple intermediate $\eta$-steps are introduced, a prevailing strategy in free energy calculations with substantially different configurational spaces~\cite{frenkel2002understanding}. Structurally, the $\eta$-ensemble becomes reminiscent of the TI with the in-between steps. However, in the $\eta$-ensemble, $\eta$-values with a finite difference are considered, whereas in TI a continuous function of the coupling constant is integrated.

The $\eta$-ensemble is performed in the bosonic sector and is therefore FSP free, but a large number of intermediate $\eta$-steps will result in a prohibitive computational cost for accurate free energy calculations for large $N$. The majority of the computational cost in each MC step comes from the evaluation of the Ewald summation. To avoid this problem, we evaluate the $\eta$-ensemble using a nonphysical interaction or artificial interaction, $\Hat{V} \rightarrow \Hat{V}_{\text{art}}$, which is computationally cheap, and any error is corrected for in a second step henceforth referred to as the $a$-ensemble. The $a$-ensemble concerns the Hamiltonian 
\begin{equation}
    \Hat{H}_a = \Hat{K} + a \Hat{V} + (1 - a) \Hat{V}_{\text{art}},
    \label{eq:a_ensemble_hamiltonian}
\end{equation}
were $a = 0$ and $a = 1$ is used for the two systems in Equation \eqref{eq:extended_ensemble_free_energy}. If the physical interaction $\Hat{V}$ and the artificial one $\Hat{V}_{\text{art}}$ are sufficiently similar, no intermediate $a$ steps are required, as no substantial energy penalty is incurred when altering $a$. This procedure accelerates the computation as the majority of the data collection is performed with a fast artificial interaction, but it does not constitute any approximation as it can simply be viewed as establishing a transitional reference system, as is common practice when performing TI~\cite{alfe1999melting,caillol2000monte,plummer2025ionization}.

The artificial interaction in question is in principle a free choice, but it should be both efficient and close to the physical one to avoid unnecessary computations in the $a$-ensemble. Working with the Coulomb interaction, a variety of cut-off based approximations have been developed, which all could be used as the artificial interaction; see review by \citeauthor{fukuda2012non}~\cite{fukuda2012non} and references therein. In this work, we have used the spherically averaged Ewald potential by \citeauthor{yakub2003efficient} (YR)~\cite{yakub2003efficient,yakub2005new} that has been successfully applied in MD~\cite{yakub2007molecular}, MC~\cite{demyanov2022one,demyanov2024N} and PIMC~\cite{filinov2020uniform,dornheim2025application}, and recently has attracted new theoretical interest~\cite{demyanov2022systematic}. By construction, the YR interaction yields energies similar to the Ewald summation, and its simple algebraic structure makes it cheap to evaluate allowing for classical MC simulation with up to $10^{6}$ particles~\cite{demyanov2022one}. The new $a$-ensemble with the YR potential as the artificial interaction has been implemented in the ISHTAR code~\cite{dornheim2024ISHTAR}, which employs the canonical adaptation~\cite{mezzacapo2007structure,dornheim2021ab} of the worm algorithm~\cite{boninsegni2006worm_a,boninsegni2006worm_b}. All reported computations have been performed using the primitive factorisation~\cite{ceperley1995path}. 

\begin{figure*}
    \centering
    \includegraphics[]{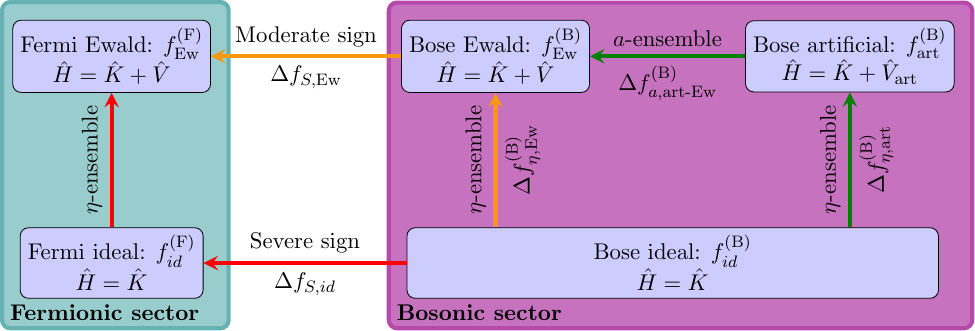}
    \caption{Schematic showing the different systems used to compute the free energies $f$ and the Hamiltonians $\Hat{H}$ of each system. Arrows indicate an ensemble or sign computation to go between systems, and labels the associated free energy change $\Delta f$. Green arrows indicate computations which are computational cheap, while orange arrows indicate moderate cost either due to a costly Ewald computation or the FSP. Red arrows are severely affected by the FSP.}
    \label{fig:schematic}
\end{figure*}

Up to this point, only the bosonic sector has been considered to avoid the FSP. To calculate the free energy with Fermi statistics $f_{1}^{\text{(F)}}$ rather than Bose statistics $f_{1}^{\text{(B)}}$ for a system with interaction 1, the sign $S_1$ in the corresponding system should be resolved~\cite{dornheim2025direct}:
\begin{equation}
    f_{1}^{\text{(F)}} - f_{1}^{\text{(B)}} = -\frac{\boltz T}{N} \log\left( S_1 \right) \equiv \Delta f_{S, 1}.
    \label{eq:sign_contribution_free_energy}
\end{equation}
Equation \eqref{eq:sign_contribution_free_energy} completes our methodology for free energy computations which is schematically shown in Figure \ref{fig:schematic} highlighting the steps and Hamiltonians.

\begin{figure}
    \centering
    \includegraphics[width=\linewidth]{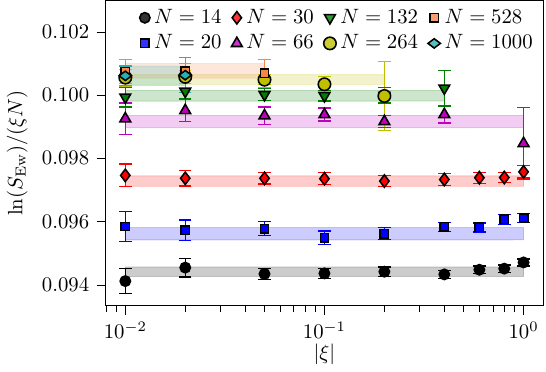}
    \caption{The average sign $S_{\text{Ew}}$ for the UEG at $r_s = 3.23$ and $\theta = 1.0$, for different system sizes $N$ and permutation weight $\xi$. Note that $\xi < 0$ and all values of $S_{\text{Ew}}$ are less than unity. The error bars, correspond to $95\%$-confidence intervals estimated from simulations with varying initial conditions. The extrapolation of the confidence interval assuming $a_S$ is independent of $\xi$ is shown in the highlighted areas. The point from which the extrapolation is performed is described in the supporting information. Good agreement with the extrapolation is demonstrated, validating the computational model for $S_{\text{Ew}}$.}
    \label{fig:sign_extrapolation}
\end{figure}

For the sign evaluation in Equation \eqref{eq:sign_contribution_free_energy}, the above-mentioned $\xi$-extrapolation was used based on the functional form:
\begin{equation}
    S(N, \xi) = e^{a_S(N, \xi)\, N \xi},
    \label{eq:sign_xi_extrap}
\end{equation}
where the primary scaling with $N$ and $\xi$ is factored out, and the remaining function $a_S(N, \xi)$ shows only small deviations from being constant. \citeauthor{dornheim2025fermionic} successfully showed that the extrapolation from $\xi = -0.2$ based on Equation \eqref{eq:sign_xi_extrap} with $a_S(N, \xi) = a_S(N)$ is highly accurate for $\theta = 1$~\cite{dornheim2025fermionic}. Figure \ref{fig:sign_extrapolation} greatly extends the validation of this extrapolation method by considering a two orders of magnitude range for $\xi$ for $N \leq 66$. Validation of the method to substantially smaller $\xi$ is crucial for modeling larger system sizes, since keeping $\xi N$ roughly constant maintains a resolvable sign. System sizes up to $N = 1000$ are investigated in Figure \ref{fig:sign_extrapolation}, and $N \geq 264$ is observed to be needed to converge the finite-size effect to within the statistical error bars. This highlights the need to model large systems to approach the thermodynamic limit.

The minor systematic error observed in the $\xi$-extrapolation with $N = 14$ is $0.3\%$ and corresponds to a $0.05\,\text{mHa}$ error in free energy. These errors are expected to decrease with the size of the system, where the permutation structure is less affected by boundary effects and self-exchanges~\cite{dornheim2019path}; this makes the generalization of the corresponding free energy difference via the $\xi$-extrapolation more straightforward. This can be seen particularly well for the more strongly coupled case of $r_s=10$ shown in the supporting information.

\begin{figure}
    \centering
    \includegraphics[width=\linewidth]{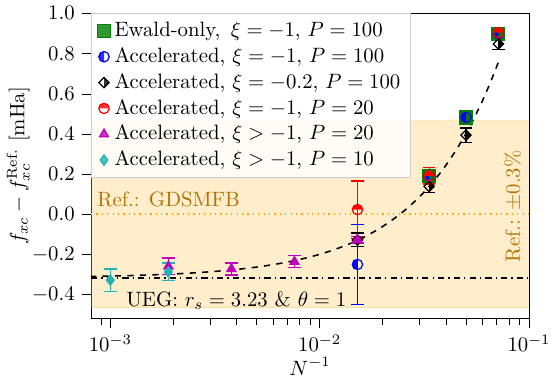}
    \caption{Finite size and finite $P$ corrected exchange correlation free energies for the UEG at $r_s = 3.23$ and $\theta = 1.0$ shown as the difference from the GDSMFB parametrisation~\cite{groth2017ab} with a value of $f_{xc}^{\text{Ref.}} = -0.15529\,\text{Ha}$ (Ref.). The remaining $N$ dependence is a fraction of a percent. The computations have been performed in a variety of ways, using the physical interaction (Ewald-only) or accelerated method (Accelerated), with ($\xi > -1$) and without ($\xi = -1$) $\xi$-extrapolation, and varying number of propagators $P$. Overlapping data points are shown when $P$ is reduced or extrapolation techniques are employed, demonstrating the correctness of the procedure. The dashed line is a fit on the form $f_{xc}(N) = c_0 + c_1 N^{-c_2}$ for $N \leq 30$, where $c_0, c_1$ and $c_2$ are fitting coefficients. The extrapolated free energy is reduced by $0.2\%$ compared to the reference and $c_2 \approx 1.3$. Error bars as described in Figure \ref{fig:sign_extrapolation}.}
    \label{fig:xc_free_energy}
\end{figure}

The nonideal contribution to the fermionic free energy is the exchange correlation free energy:
\begin{align}
    f_{xc} &= f_{\text{Ew}}^{(\text{F})} - f_{id}^{(\text{F})}\\
    &= \Delta f_{\eta, \text{art}}^{(\text{B})} + \Delta f_{a, \text{art-Ew}}^{(\text{B})} + \left(\Delta f_{S, \text{Ew}} - \Delta f_{S, id}\right),\nonumber
\end{align}
which in our accelerated scheme (second line) has three distinct contributions. The contribution of the $\eta$-ensemble with the artificial interaction $\Delta f_{\eta, \text{art}}^{(\text{B})}$, the correction from the $a$-ensemble $\Delta f_{a, \text{art-Ew}}^{(\text{B})}$ and the difference between the sign contribution for the interacting and noninteracting system $\Delta f_{S, \text{Ew}} - \Delta f_{S, id}$. In the standard Ewald-only approach, the first two contributions are given by a single term $\Delta f_{\eta, \text{Ew}}^{(\text{B})} = \Delta f_{\eta, \text{art}}^{(\text{B})} + \Delta f_{a, \text{art-Ew}}^{(\text{B})}$. The origin of each term is also shown in Figure \ref{fig:schematic}.

As both a conceptual and practical validation of the acceleration method, Figure \ref{fig:xc_free_energy} shows the exchange correlation free energy computed both via the standard Ewald-only method and our accelerated scheme for $N \leq 30$. The results cannot be distinguished from each other on the scale of Figure \ref{fig:xc_free_energy}, and any deviation lies within the statistical error margins. As the system size increases, the accelerated method can perform up to 18 times as many Monte Carlo steps as the standard method in a given time; see the supporting information for additional information. The additional Monte Carlo samples reduce the statistical error but more crucially allow us to investigate larger system sizes. 

In Figure \ref{fig:xc_free_energy}, the exchange correlation free energy computations are scaled up to 1000 electrons using the accelerated method. To limit computational expense, a reduced number of imaginary time slices $P$ is used to factorise the density matrix for large $N$. The finite $P$ error has been systematically investigated for smaller $N$ with $P$s between 8 and 200 as demonstrated in the supporting information. Empirically, we find that the corresponding $P$-correction that connects a finite $P$ to the limit of $P^{-1} \rightarrow 0$ is independent of $N$, reflecting the local nature of the factorization error, which is ultimately due to the quantum delocalization of individual particles. The correction has been applied in Figure \ref{fig:xc_free_energy}. As a further validation of the finite-$P$ correction, duplicate data points are shown when $P$ is reduced and the results are always within the statistical error.

To further reduce the size dependence of the free energy, the results in Figure \ref{fig:xc_free_energy} have been finite-size corrected using the method introduced by \citeauthor{groth2017ab}~\cite{groth2017ab} (see further details in the supporting information). The finite-size correction is highly efficient and at the investigated condition removes $93\%$ of the finite-size effect already at the smallest system used, resulting in a remaining finite-size error of the order of $1\,\text{mHa}$ per electron. In Figure \ref{fig:xc_free_energy}, it is shown that the surviving size-dependent error scales roughly linearly with $N^{-1}$ (dashed black), and for the largest systems investigated, this error is expected to be one hundredth of a mHa. The results are within $0.3\%$ of the GDSMFB parametrisation computed by the adiabatic connection formula~\cite{groth2017ab}, well within the expected error margins of their parametrisation. In conclusion, the finite size correction method by \citeauthor{groth2017ab} is highly efficient and virtually any remaining finite size error has been eliminated by reaching system sizes with 1000 electrons, now numerically feasible with our accelerated technique for free energy calculations.

\begin{figure}[t]
    \centering
    \includegraphics[width=\linewidth]{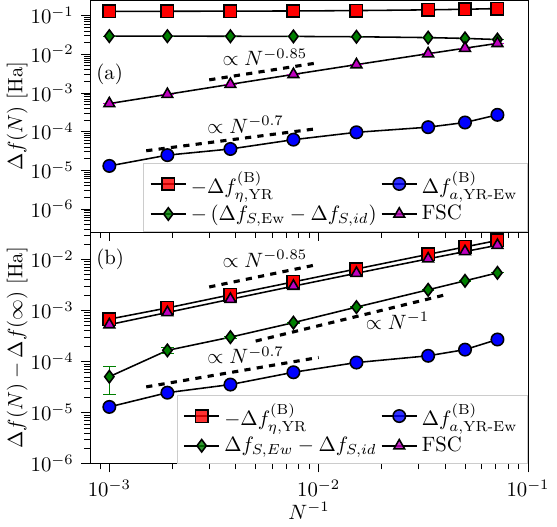}
    \caption{The size (a) and scaling (b) of the $N$ dependent free energy for the UEG at $r_s = 3.23$ and $\theta = 1.0$. The correction term $\Delta f_{a}^{\text{YR-Ew}}$ and FSC vanish as $N \rightarrow \infty$ while the contribution from the $\eta$-ensemble and the sign converges to a finite value. The subtraction of the infinite system size contribution $N = \infty$ in (b) was facilitated by a fit $\Delta f = d_0 + d_1 N^{-d_2}$ for $N \leq 66$. The interaction components are seen to converge sub-linearly ($d_2 < 1$). Error bars based on statistical error as described in Figure \ref{fig:sign_extrapolation}.}
    \label{fig:free_energy_componets}
\end{figure}

The magnitude of each of the contributions to the exchange correlation free energy is shown in Figure \ref{fig:free_energy_componets}(a). The dominant contribution under the investigated condition is the interaction contribution from the $\eta$-ensemble ($\Delta f_{\eta,\text{YR}}^{(\text{B})}$) followed by the sign contribution ($\Delta f_{S, \text{Ew}} - \Delta f_{S, id}$). The correction for using the artificial interaction in the $\eta$-ensemble ($\Delta f_{a, \text{YR-Ew}}^{(\text{B})}$) is three orders of magnitude smaller than the overall contribution of interactions. This highlights the efficiency of the YR interaction in mimicking the full Ewald summation with respect to energy, even if some artefacts are present for spatially resolved quantities~\cite{dornheim2025application}. Furthermore, the magnitude of the correction vanishes with increasing system size, as the YR potentials tend to the Coulomb form. For the $r_s = 10$ system, the picture is broadly the same, but the interaction contribution is even more dominant for this strongly interacting case.

The approach to the thermodynamic limit for the three contributions to the free energy is highlighted in Figure \ref{fig:free_energy_componets}(b) by subtracting the (fitted) thermodynamic limit. The size of the finite $N$ errors generally follows the magnitude of each respective term. The finite size error in the sign, which is by far the hardest contribution to compute in practice, is seen to scale linearly with particle number; this might be exploited for further extrapolation and optimization in future works. The two interaction contributions scale sublinearly, with approximate exponents of $0.7$ and $0.85$, respectively. However, these two exponents are not universal as they increase for the $r_s = 10$ conditions. In the supporting material, the sublinear scaling is discussed in terms of the finite-size correction model. For the simulation with $N = 1000$, all finite-size errors are below $1\,\text{mHa}$ and the chemical accuracy is reached even without any finite-size correction procedure.

To summarise, we have introduced and exemplified the use of an accelerated method for free energy estimation based on \textit{ab initio} PIMC. The method accelerates the computation in two primary ways. First, an intermediate ``artificial'' reference system is introduced in which interactions are numerically evaluated more efficiently. The majority of interaction effects can be captured in this artificial system, and any remaining error can be corrected by the introduced $a$-ensemble which in our work only required a single computation with the numerically more costly physical interaction. In this work, the use of the artificial interaction reduced the computation effort by a factor of up to 18 for the interaction contribution. Second, a $\xi$-extrapolation methodology is employed to resolve the sign for larger system sizes that are otherwise prevented by the fermionic sign problem. This extrapolation was shown to be accurate to $0.3\%$ over two orders of magnitude in $\xi$ for $\theta = 1$. The generality of the procedure was demonstrated by considering two different density conditions.

Accelerating the calculation of free energies paves the way for scaling up computations to remove the final systematic error -- the finite size effects -- at warm dense matter conditions. The presented method can be combined with other acceleration techniques to consider even large systems, e.g., employing GPUs~\cite{thorpe2025acceleration}, hierarchical energy evaluation~\cite{muller2023fast}, and contraction schemes~\cite{john2016quantum}. High-precision free energy estimates for the UEG open for the possibility to explore a potential spin phase transition at finite temperature, which have been intensely studied in the ground state~\cite{ceperley1980ground,loos2016uniform}. Future work might also explore the long-wavelength physics with the presented method via the density stiffness theorem~\cite{giuliani2008quantum,dornheim2023energy}, which relates the static linear and non-linear density response to free energy differences. In this regard, the simulation of large systems is crucial to study the optical limit of $\vec{k} \rightarrow 0$, where the minimum wavenumber $|\vec{k}| = 2\pi/L$ is determined by the box length $L$. Lastly, the present study focuses on the UEG, but it is straightforward to apply our methodology to light elements such as hydrogen and beryllium~\cite{dornheim2025unraveling} to inform planetary and inertial confinement fusion modeling.
Moreover, our approach can easily be applied to the simulation of inhomogeneous systems such as electrons in a fixed ionic configuration, which might be of great value for the benchmarking of DFT and potentially even for the data-driven construction of novel exchange correlation functionals~\cite{PhysRevLett.127.126403}.

\begin{acknowledgement}
This work was partially supported by the Center for Advanced Systems Understanding (CASUS), financed by Germany’s Federal Ministry of Education and Research and the Saxon state government out of the State budget approved by the Saxon State Parliament. Further support is acknowledged for the CASUS Open Project \textit{Guiding dielectric theories with ab initio quantum Monte Carlo simulations: from the strongly coupled electron liquid to warm dense matter}. This work has received funding from the European Research Council (ERC) under the European Union’s Horizon 2022 research and innovation programme (Grant agreement No. 101076233, "PREXTREME"). 
Views and opinions expressed are however those of the authors only and do not necessarily reflect those of the European Union or the European Research Council Executive Agency. Neither the European Union nor the granting authority can be held responsible for them. Computations were performed on a Bull Cluster at the Center for Information Services and High-Performance Computing (ZIH) at Technische Universit\"at Dresden and at the Norddeutscher Verbund f\"ur Hoch- und H\"ochstleistungsrechnen (HLRN) under grant mvp00024.
\end{acknowledgement}

\begin{suppinfo}
Additional details for Monte Carlo updates in the $a$-ensemble, numerical parameters, discussion on acceleration in the $\eta$-ensemble, finite size and finite $P$ correction procedures, and the case study of the UEG at the density $r_s = 10$ are shown in the supporting information.
\end{suppinfo}

\iftrue
\onecolumn
\iftrue
\subsection*{\textbf{Supporting Information for ``Accelerated free energy estimation in \emph{ab initio} path integral Monte Carlo simulations''}}
\else
\documentclass[journal=jpclcd,manuscript=article]{achemso}

\usepackage[version=3]{mhchem} 
\usepackage{lmodern}
\usepackage{color}
\usepackage[dvipsnames,svgnames,table]{xcolor}
\usepackage[colorlinks=true,linkcolor=blue,urlcolor=blue,citecolor=blue]{hyperref}
\usepackage{graphicx}
\usepackage{dcolumn}
\usepackage{array}
\usepackage{bm}
\usepackage{subfigure}
\usepackage{amssymb}
\usepackage{multirow}
\usepackage{tabularx}
\usepackage{amsmath}
\usepackage{makecell}


\newcommand{\deri}[2]{\frac{\partial #1}{\partial #2}}
\newcommand{\derii}[2]{\frac{\partial^2 #1}{\partial #2^2}}
\renewcommand{\vec}[1]{\boldsymbol{#1}}
\newcommand{\mat}[1]{\mathbf{#1}}
\DeclareMathSymbol{\shortminus}{\mathbin}{AMSa}{"39}
\newcommand{\erfc}{\text{erfc}}
\newcommand{\boltz}{k_{\text{B}}}
\newcommand{\bohr}{a_{\text{B}}}
\newcommand{\order}[1]{\mathcal{O}\left(#1\right)}
\DeclareMathOperator*{\argmax}{arg\,max}
\DeclareMathOperator*{\argmin}{arg\,min}

\iftrue
    \newcommand{\todo}[1]{{\color{red}[#1]}} 
    \newcommand{\question}[1]{{\color{blue}[#1]}}
\else
    \newcommand{\todo}[1]{}               
    \newcommand{\question}[1]{}
\fi

\author{Pontus Svensson}
\email{p.svensson@hzdr.de}
\affiliation{Center for Advanced Systems Understanding (CASUS), D-02826 G\"orlitz, Germany}
\alsoaffiliation{Helmholtz-Zentrum Dresden-Rossendorf (HZDR), D-01328 Dresden, Germany}

\author{Fotios Kalkavouras}
\affiliation{Space and Plasma Physics, Royal Institute of Technology (KTH), Stockholm, SE-100 44, Sweden}

\author{Uwe Hernandez Acosta}
\affiliation{Center for Advanced Systems Understanding (CASUS), D-02826 G\"orlitz, Germany}
\alsoaffiliation{Helmholtz-Zentrum Dresden-Rossendorf (HZDR), D-01328 Dresden, Germany}

\author{Zhandos~A.~Moldabekov}
\affiliation{Center for Advanced Systems Understanding (CASUS), D-02826 G\"orlitz, Germany}
\alsoaffiliation{Helmholtz-Zentrum Dresden-Rossendorf (HZDR), D-01328 Dresden, Germany}

\author{Panagiotis Tolias}
\affiliation{Space and Plasma Physics, Royal Institute of Technology (KTH), Stockholm, SE-100 44, Sweden}

\author{Jan Vorberger}
\affiliation{Helmholtz-Zentrum Dresden-Rossendorf (HZDR), D-01328 Dresden, Germany}

\author{Tobias Dornheim}
\affiliation{Center for Advanced Systems Understanding (CASUS), D-02826 G\"orlitz, Germany}
\alsoaffiliation{Helmholtz-Zentrum Dresden-Rossendorf (HZDR), D-01328 Dresden, Germany}

\title{Supporting Information for ``Accelerated free energy estimation in \emph{ab initio} path integral Monte Carlo simulations''}

\keywords{Free energy, PIMC, Warm dense matter, Uniform electron gas}

\begin{document}
\singlespacing
\fi

\newcommand{\mainrefschematic}{1}
\newcommand{\mainrefsignextrapolation}{2}
\newcommand{\mainrefxcfreeenergy}{3}
\newcommand{\mainreffreeenergycomponets}{4}

\newcommand{\mainrefextendedensemble}{(1)}
\newcommand{\mainrefaensemblehamiltonian}{(3)}

\setcounter{equation}{0}
\renewcommand{\theequation}{S\arabic{equation}}

\setcounter{figure}{0}
\renewcommand{\thefigure}{S\arabic{figure}}

\setcounter{table}{0}
\renewcommand{\thetable}{S\Roman{table}}

\section{Extended ensemble and corrections via the $a$-ensemble}

The extended ensemble method~\cite{dornheim2025direct} is used to calculate the contribution of interactions to the free energy (in the bosonic sector) in two steps. First, the free-energy difference between the interacting and ideal bosonic systems is calculated using the $\eta$-ensemble. In particular, for a large number of particles the acceptance probability is reduced for the move which alters $\eta$ as the likelihood of finding two electrons in close proximity increases in the ideal system, configurations which are strongly suppressed in the interacting one~\cite{dornheim2025eta}. Therefore, to maintain an ergodic exploration of the extended ensemble, the $\eta$-ensemble is subdivided into $N_{\eta}$ steps and the free energy difference is computed for a set of $\eta$-values, i.e. $\left\{ \eta_i \right\}_{i = 1}^{N_{\eta} + 1}$, where $\eta_1 = 1$, $\eta_{N_{\eta} + 1} = 0$ and $\eta_{i + 1} < \eta_{i}$. To accelerate this process, these computations are carried out with an artificial interaction which mimics our target interaction -- the Ewald summation in this case -- but is numerically less expensive to evaluate. Here, the spherically averaged Ewald interaction by Yakub and Ronchi (YR)~\cite{yakub2003efficient,yakub2005new} was used. The details of the moves in the $\eta$-ensemble were given by \citeauthor{dornheim2024eta}~\cite{dornheim2024eta}. Second, the $a$-ensembel is used to correct for the use of the artificial interaction in the $\eta$-ensemble by considering the Hamiltonian in Equation \mainrefaensemblehamiltonian{} of the main text. In principle, the $a$-ensemble can be subdivided into $N_a$ computations with a set of $a$-values $\left\{ a_i \right\}_{i = 1}^{N_{a} + 1}$, where $a_1 = 1$, $a_{N_{\eta} + 1} = 0$ and $a_{i+1} < a_{i}$. However, if the artificial interaction is sufficiently similar to the physical one, a single step $N_a = 1$ is sufficient to maintain ergodicity, as has been the case in this work. This is how the method accelerates the computations, as a reduced number of calculations with the more expensive Ewald interactions is needed.

The moves to switch $a$-values are constructed analogously to the moves in the $\eta$-ensemble with the Metropolis-Hastings~\cite{metropolis1953equation,hastings1970monte} acceptance probabilities
\begin{subequations}
\begin{equation}
    A(a_i\to a_{i+1}) = \textnormal{min}\left\{1, \frac{1}{c_{a_i}}\textnormal{exp}\left(\epsilon\left\{ (a_i - a_{i+1}) V(\vec{X}) + (a_{i+1}-a_i) V_\textnormal{art}(\vec{X}) \right\}\right)\right\},
\end{equation}
and 
\begin{eqnarray}
    A(a_{i+1}\to a_{i}) = \textnormal{min}\left\{1, {c_{a_i}}\vphantom{\frac{1}{c_{a_i}}}\textnormal{exp}\left(\epsilon\left\{ (a_{i+1} - a_{i}) V(\vec{X}) + (a_{i}-a_{i+1}) V_\textnormal{art}(\vec{X}) \right\}\right)\right\},
\end{eqnarray}
\end{subequations}
where $\vec{X}$ is the path configuration, $c_{a_i}$ corresponds to $c$ in Equation \mainrefextendedensemble{}, $\epsilon = \beta / P$, and $P$ is the number of factorisations of the density matrix. The move only modifies the $a$-value, and not the path configurations $\vec{X}$. Furthermore, the move is applied in both the diagonal and off-diagonal sectors of the worm algorithm~\cite{boninsegni2006worm_a,boninsegni2006worm_b}. 

The resulting free energy computed from a series of PIMC calculations is $f_{\text{Ew}}^{(\text{F})} = f_{id}^{(\text{B})} + \Delta f_{\eta,\text{Ew}}^{(\text{B})} + \Delta f_{S,\text{Ew}}$ or using the acceleration method $f_{\text{Ew}}^{(\text{F})} = f_{id}^{(\text{B})} + \Delta f_{\eta,\text{art}}^{(\text{B})} + \Delta f_{a,\text{art-Ew}}^{(\text{B})} + \Delta f_{S,\text{Ew}}$, where
\begin{subequations}
\begingroup
\allowdisplaybreaks
\begin{align}
    \Delta f_{\eta,\text{Ew}}^{(\text{B})} &= -\frac{1}{\beta N} \sum_{i = 1}^{N_{\eta}} \log\left[ \frac{r_{\eta_i,\eta_{i+1}}^{\text{Ew}}}{c_{\eta_{i}}} \right],\\
    \Delta f_{\eta,\text{art}}^{(\text{B})} &= -\frac{1}{\beta N} \sum_{i = 1}^{N_{\eta}} \log\left[ \frac{r_{\eta_i,\eta_{i+1}}^{\text{art}}}{c_{\eta_{i}}} \right],\\
    \Delta f_{a,\text{art-Ew}}^{(\text{B})} &= -\frac{1}{\beta N} \sum_{i = 1}^{N_{a}} \log\left[ \frac{r_{a_i,a_{i+1}}^{\text{art-Ew}}}{c_{a_{i}}} \right],
\end{align}
\endgroup
and 
\begin{align}
    \Delta f_{S,\text{Ew}} &= -\frac{1}{\beta N} \log\left[ S_{\text{Ew}} \right], 
\end{align}
\end{subequations}
and see Figure \mainrefschematic{} in the main text for the origin of each term. In the above, $c_{\eta_i}$ corresponds to $c$ in Equation \mainrefextendedensemble{} in the main text. The ratios of samples in each subsystem in the extended ensemble is $r_{\eta_{i}, \eta_{i+1}}^{\text{Ew}}$, $r_{\eta_{i}, \eta_{i+1}}^{\text{art}}$ and $r_{a_{i}, a_{i+1}}^{\text{art-Ew}}$ in both the $\eta$ and $a$-ensembles for the Ewald (Ew) and artificial (art) interactions. The implementation of the $\eta$-ensemble was benchmarked by \citeauthor{dornheim2024eta}~\cite{dornheim2024eta} and the implementation of the $a$-ensemble is confirmed by comparing the free energies between the Ewald-only method and our acceleration method. The different schemes agree within the statistical errors; see Figure \mainrefxcfreeenergy{} in the main text. 

\section{Numerical parameters for simulations}

\begin{table}[]
    \centering
    \caption{Summary of computational parameters. The $\xi$-column describe the $\xi$ value used for the extrapolation in Figures \mainrefxcfreeenergy{} and \mainreffreeenergycomponets{} in the main text, and Figure \ref{fig:free_energy_rs_10}. The $P$-column describe the number of imaginary time slices used for the computation in Figures \mainrefsignextrapolation{} and \mainreffreeenergycomponets{} in the main text, and Figure \ref{fig:free_energy_rs_10}. The three final columns gives the details of the $\eta$-ensemble computation. Note that an $a$-ensemble simulation has been performed in each  case.}
    \label{tab:numerical_parameters}
    \rowcolors{2}{gray!15}{white}
    \newcolumntype{C}{>{\centering\let\newline\\\arraybackslash\hspace{0pt}}m{6.2cm}}
    \begin{tabular}{|c|cccCC|}
        \hline
        \multicolumn{6}{|c|}{\cellcolor{gray!40}$r_s = 3.23$ \& $\theta = 1.0$}\\\hline 
         $N$ & $\xi$ & $P$ & $N_\eta$ & $\left\{\eta_i \right\}_{i = 1}^{N_{\eta} + 1}$ & $\left\{ c_{\eta_i} \right\}_{i = 1}^{N_{\eta}}$ \\\hline
         14  & 0.2 & 100 & 4 & \{0.0, 0.01, 0.1, 0.5, 1.0\} & \{1, 5e-1, 7e-3, 10e-4\}\\
         20  & 0.2 & 100 & 4 & \{0.0, 0.01, 0.1, 0.5, 1.0\} & \{1, 2e-1, 2e-3, 1e-4\}\\
         30  & 0.2 & 100 & 4 & \{0.0, 0.01, 0.1, 0.5, 1.0\} & \{1, 1e-1, 1e-4, 1e-6\}\\
         66  & 0.2 & 100 & 7 & \{0.0, 0.01, 0.1, 0.2, 0.4, 0.6, 0.8, 1.0\} & \{1, 2e-1, 3e-2, 2e-4, 6e-5, 1.5e-5, 3e-6\}\\
         132 & 0.1 & 20  & 7 & \{0.0, 0.01, 0.1, 0.2, 0.4, 0.6, 0.8, 1.0\} & \{1, 1e-1, 1e-2, 1e-7, 8e-10, 2e-10, 1e-11\}\\
         264 & 0.05 & 20 & 12 & \{0.0, 0.01, 0.05, 0.1, 0.2, 0.3, 0.4, 0.5, 0.6, 0.7, 0.8, 0.9, 1.0\} & \{1, 1, 6e-2, 8e-6, 4e-7, 1e-8, 1e-8, 2e-10, 1e-10, 3e-11, 1e-11, 5e-12\}\\
         528 & 0.02 & 20 & 23 &  \{0.0, 0.01, 0.025, 0.05, 0.075, 0.1, 0.15, 0.2, 0.25, 0.3, 0.35, 0.4, 0.45, 0.5, 0.55, 0.6, 0.65, 0.7, 0.75, 0.8, 0.85, 0.9, 0.95, 1.0\} &  \{1, 1, 5e-1, 1e-2, 1e-2, 1e-5, 1e-6, 1e-6, 5e-8, 2e-8, 1e-8, 1e-9, 1e-9, 1e-9, 1e-10, 1e-10, 1e-10, 1e-10, 1e-11, 1e-11, 1e-11, 1e-11, 1e-11\}\\
         1000 & 0.01 & 10 & 23 & \{0.0, 0.01, 0.025, 0.05, 0.075, 0.1, 0.15, 0.2, 0.25, 0.3, 0.35, 0.4, 0.45, 0.5, 0.55, 0.6, 0.65, 0.7, 0.75, 0.8, 0.85, 0.9, 0.95, 1.0\} & \{1, 1, 1e-1, 2e-3, 1e-3, 1e-9, 1e-11, 2e-13, 1e-14, 1e-15, 1e-16, 1e-17, 1e-17, 1e-18, 1e-18, 1e-19, 1e-19, 1e-20, 1e-20, 1e-20, 1e-21, 1e-21, 1e-21\}\\\hline
         \multicolumn{6}{|c|}{\cellcolor{gray!40}$r_s = 10.0$ \& $\theta = 1.0$}\\\hline 
         $N$ & $\xi$ & $P$ & $N_\eta$ & $\left\{\eta_i \right\}_{i = 1}^{N_{\eta} + 1}$ & $\left\{ c_{\eta_i} \right\}_{i = 1}^{N_{\eta}}$ \\\hline
         14  & 0.2 & 100 & 5 & \{0.0, 0.01, 0.1, 0.2, 0.5, 1.0\} & \{1.0, 1e-1, 1e-2, 1e-6, 1e-11\}\\
         30  & 0.2 & 100 & 5 & \{0.0, 0.01, 0.1, 0.2, 0.5, 1.0\} & \{1.0, 1e-2, 1e-3, 5e-13, 1e-24\}\\
         66  & 0.2 & 100 & 7 & \{0.0, 0.01, 0.1, 0.2, 0.4, 0.6, 0.8, 1.0\} & \{1.0, 1e-4, 1e-7, 5e-18, 1e-20, 1e-21, 5e-22\}\\
         132  & 0.1 & 20 & 9 & \{0.0, 0.02, 0.05, 0.1, 0.2, 0.3, 0.4, 0.6, 0.8, 1.0\} & \{1e-1, 1e-2, 1e-5, 1e-14, 1e-16, 1e-18, 2e-39, 1e-41, 1e-43\}\\
         264  & 0.1 & 20 & 11 & \{0.0, 0.02, 0.05, 0.1, 0.15, 0.2, 0.3, 0.4, 0.5, 0.6, 0.8, 1.0\} & \{1e-1, 1e-4, 1e-10, 1e-13, 1e-15, 1e-33, 1e-36, 1e-38, 1e-39, 1e-82, 1e-85\}\\
         528  & 0.1 & 20 & 16 & \{0.0, 0.01, 0.02, 0.05, 0.075, 0.1, 0.15, 0.2, 0.25, 0.3, 0.4, 0.5, 0.6, 0.7, 0.8, 0.9, 1.0\} & \{1, 1e-1, 1e-8, 1e-9, 1e-11, 1e-26, 1e-29, 1e-32, 1e-34, 1e-72, 1e-76, 1e-79, 1e-81, 1e-83, 1e-85, 1e-86\}\\
         1000  & 0.02 & 20 & 25 & \{0.0, 0.001, 0.002, 0.005, 0.01, 0.02, 0.03, 0.05, 0.075, 0.1, 0.125, 0.15, 0.175, 0.2, 0.25, 0.3, 0.35, 0.4, 0.45, 0.5, 0.55, 0.6, 0.7, 0.8, 0.9, 1.0\} & \{1, 1, 1, 1, 1e-2, 1e-4, 1e-11, 1e-18, 1e-21, 1e-24, 1e-26, 1e-27, 1e-29, 1e-61, 1e-64, 1e-67, 1e-69, 1e-71, 1e-73, 1e-74, 1e-75, 1e-154, 1e-158, 1e-161, 1e-163\}\\\hline
    \end{tabular}
\end{table}

Some additional numerical parameters used for the PIMC simulations are provided in Table \ref{tab:numerical_parameters}, including the $\xi$-points used for extrapolation and the number of time slices $P$ used to represent the density matrix. In addition, the parameters used for the $\eta$-ensemble are shown. The number of subdivisions $N_{\eta}$ increases with $N$, and the $\eta_i$ grid is nonuniform, as the structural properties of the system change more rapidly with respect to $\eta$ when approaching the noninteracting limit. This is particularly evident for $N = 1000$ and $r_s = 10.0$ where the last step in $\eta$ is one hundred times larger than the first, while retaining roughly the same acceptance probability for the $\eta$-move. The exact choice of $c_\eta$ does not influence the result~\cite{dornheim2025eta}, but for algorithmic efficiency the number of samples in the two partition functions should be approximately equal. Therefore, $\ln c_\eta \approx - \beta N (f_2 - f_1)$, where $f_1$ and $f_2$ are the free energy per particle in the two systems in question which are \textit{a priori} unknown. The coefficients $c_\eta$ in Table \ref{tab:numerical_parameters} were obtained by scanning $c_\eta$ and optimising the acceptance probability. However, the results agree well with the mentioned estimate, even if the free energies are approximated by a classical parametrisation~\cite{plummer2025ionization}, except for small $\eta$ where the quantum statistics are more prevalent. 

In simulations which utilised the Ewald interaction, the Ewald parameters were optimised such that the energy of the system was converged to six significant digits, using the single image convention in real-space and using a maximal $k$-vector component of $8\pi/L$ ($L$ is the box length) in reciprocal space.

\section{Computational speedup in $\eta$-ensemble}
Empirically, we find that the computational cost to perform a Monte Carlo step in ISHTAR -- averaged over all types of steps -- is approximately:  
\begin{equation}
    \text{Computational cost}(N, P) = C_0 P + C_1^{\text{Int.}} PN,
    \label{eq:cost_model}
\end{equation}
where $C_0$ refers to computations in the update step and $C_1^{\text{Int.}}$ relates to the computational cost of evaluating the interaction. The latter scales as $PN$ as an order $N$ computation is required to evaluate the potential for each time slice. The speedup in the $\eta$-ensemble is achieved as $C_1^{\text{Int.}}$ for the YR interaction is substantially smaller than that for the Ewald interaction.

\begin{figure}
    \centering
    \includegraphics[width=0.5\linewidth]{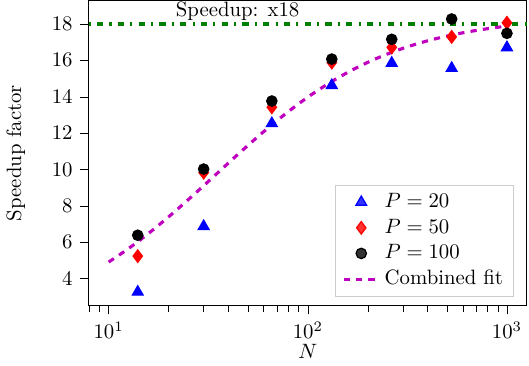}
    \caption{The computational speedup from using the spherically averaged Ewald interaction rather than the full Ewald interaction in the $\eta$-ensemble for varying number of $P$. A fit for the speedup is based on the computational model in Equation \eqref{eq:cost_model}, which predicts a maximal speedup of approximately $18$ times.}
    \label{fig:speedup}
\end{figure}

The computational acceleration was tested by performing computations with both interaction types for a subset of computations needed to evaluate the $\eta$-ensemble, and the results are shown in Figure \ref{fig:speedup}. The speedup shows only minor variations with respect to $P$ and tends toward a constant for large $N$, two properties that are well explained by the model in Equation \eqref{eq:cost_model}. For smaller $P$, numerical overhead not included in Equation \eqref{eq:cost_model} becomes more appreciable. A combined fit over $N$ and $P$ based on the ratio of Equation \eqref{eq:cost_model} for the YR and Ewald interactions is shown to appropriately represent the data. For the larger system sizes investigated, we observe an acceleration of up to 18 times. The exact numerical speedup will depend on the simulation configurations and implementation details, but the results shown here are representative of the computations in the main text.

\section{Finite size corrections (FSC)}
The finite-size corrections (FSC) applied follow the methodology given in the supplementary material of \citeauthor{groth2017ab}~\cite{groth2017ab}. The finite-size error for the exchange correlation free energy at $r_s$ and $\theta$ is given by
\begin{equation}
    \Delta f_{xc}(r_s, \theta) = \frac{1}{r_s^2} \int_{0}^{r_s} d\Bar{r}_s\; \Bar{r}_s \Delta v(\Bar{r}_s, \theta),
    \label{eq:FSC_fxc}
\end{equation}
where $\Delta v(r_s, \theta; N)$ is the finite size error on the interaction energy. The major contribution to $\Delta v$ is the discretisation error of the interaction integral imposed by the box and not the errors on the structure factor $S(\vec{k})$ itself~\cite{dornheim2016ab}. Therefore, the finite-size correction is approximated by
\begin{equation}
    \Delta v(r_s, \theta) \approx \frac{1}{2} \int \frac{d\vec{k}}{\left(2\pi\right)^{3}} \Tilde{v}_{\vec{k}} \left( \Bar{S}(\vec{k}) - 1 \right) - \left( \frac{1}{2L^{3}} \sum_{\vec{G} \neq 0} \Tilde{v}_{\vec{G}} \left( \Bar{S}(\vec{G}) - 1 \right)  + \frac{\xi_{M}}{2} \right),
    \label{eq:FSC_v}
\end{equation}
where $\Tilde{v}_{\vec{k}} = 4\pi/\vec{k}^2$ is the Fourier transformed Coulomb interaction, $L$ is the side length of the box, $\vec{G} = 2\pi \vec{n} / L$ where $\vec{n} \in \mathbb{Z}^3$, and $\xi_{M}$ is the Madelung constant. As an approximation, the static structure factor is taken from a dielectric theory, commonly the random phase approximation (RPA) $\Bar{S}(\vec{k}) = S_{\text{RPA}}(\vec{k})$. Malone implemented this procedure in \texttt{uegpy}\bibnote{See \url{https://github.com/fdmalone/uegpy}} which has been successful in removing most finite-size errors. However, for large numbers of particles this implementation suffers from some stability issues. Therefore, the procedure has been reimplemented with the classical STLS scheme~\cite{singwi1968electron} as the underlying dielectric theory, that is, $\Bar{S}(\vec{k}) = S_{\text{STLS}}(\vec{k})$.

\begin{figure*}[t]
    \centering
    \includegraphics[width=1.0\linewidth]{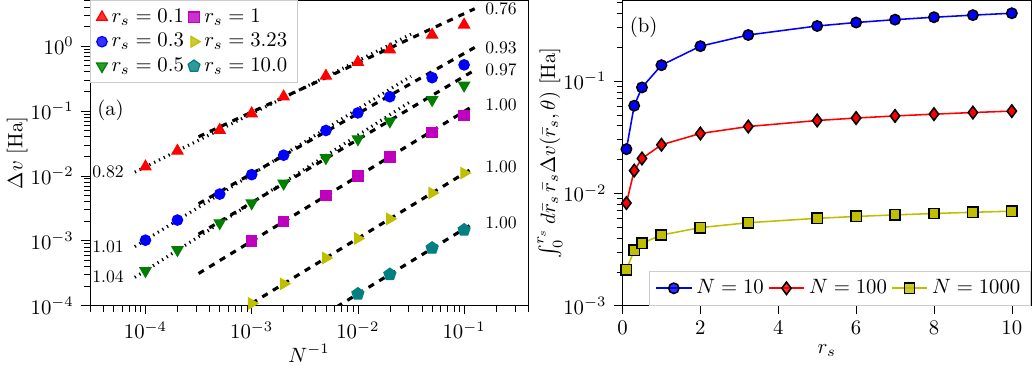}
    \caption{(a) The finite size correction of the interaction energy computed from Equation \eqref{eq:FSC_v} at $\theta = 1$ and $r_s$ between $0.1$ and $10.0$. The lines are scaling fits on the form $\Delta v = v_0 N^{-a}$, where $v_0$ and $a$ are fitting coefficients. Fits for the intervals $50 \leq N \leq 1000$ (dashed) and $1000 \leq N \leq 10000$ (dotted) are shown with their respective $a$ values to the right and left of the lines, respectively. (b) The integral in Equation \eqref{eq:FSC_fxc} is evaluated for three different particle numbers.}
    \label{fig:FSC_scaling}
\end{figure*}

To consider the $N$-scaling of $\Delta f_{xc}$, the scaling of $\Delta v$ must first be established. For the two conditions under primary investigation here, $r_s = 3.23$ and $r_s = 10.0$, the finite-size error of the interaction energy is seen to scale linearly in Figure \ref{fig:FSC_scaling}(a). However, when considering $r_s < 1.0$, we observe a sublinear scaling for particle numbers in the range $50 \leq N \leq 1000$; see Figure \ref{fig:FSC_scaling}(a). For a fixed $\theta$, small $r_s$ corresponds to the weak coupling limit as the classical coupling constant $\Gamma_{\text{cl.}}$ which characterises a classical plasma scale as $\Gamma_{\text{cl.}} \propto r_s / \theta$. The Debye length $\lambda_D \propto r_s \Gamma_{\text{cl.}}^{-1/2}$ which is the typical scale length of weakly coupled plasmas grows large compared to the inter-particle separation for small $\Gamma_{\text{cl.}}$, and large numbers of particles must be considered in the modeling. Therefore, we observe an alteration of the $N$-scaling for $r_s < 1.0$ when considering $N > 1000$. See \citeauthor{caillol2010accurate}~\cite{caillol2010accurate} and \citeauthor{demyanov2024N}~\cite{demyanov2024N} for further discussion of sublinear scaling in classical MC.

The integral in Equation \eqref{eq:FSC_fxc} accumulates the finite size error of the interaction energy for $r_s$ smaller than the target value. As shown in Figure \ref{fig:FSC_scaling}(b), a considerable fraction of the integral comes from the region $r_s < 1.0$ where the sublinear scaling is observed for the particle numbers relevant to the main text. In this manner, the sublinear scaling of the energy in the weakly coupled system propagates to the free energy at higher coupling. Within the FSC model and $50 \leq N \leq 1000$, we observe exponents between $0.58$ and $0.89$ for $\theta = 1.0$ and $r_s$ in the range $0.1$ and $10.0$. Note that one of the end points of the $\eta$-ensemble is the noninteracting limit ($\eta = 0$), and the above reasoning can be translated to the PIMC simulations.  

\section{Finite number of propagators errors and corrections (FPC)}\label{sec:FPC}
\begin{figure*}[t]
    \centering
    \includegraphics[width=\linewidth]{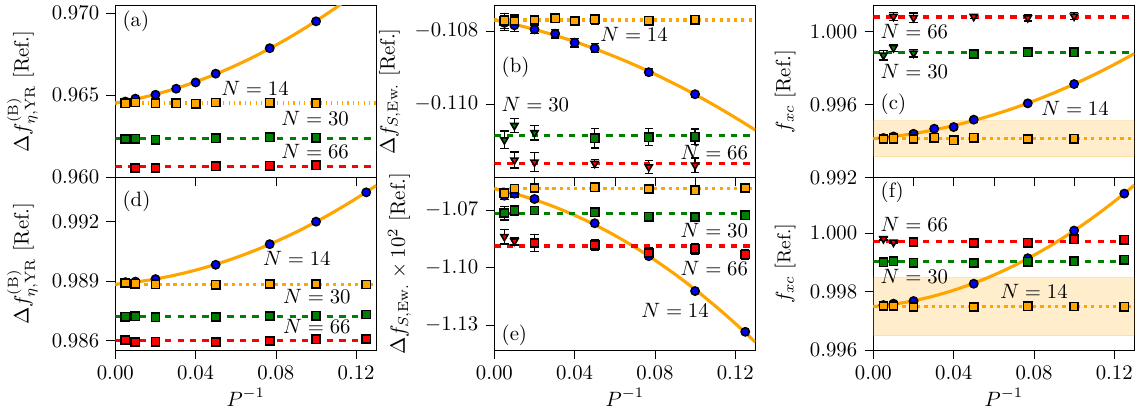}
    \caption{Free energies computed with a varying number of propagators $P$ in units of the GDSMFB parametrisation (Ref.)~\cite{groth2017ab}. Results for UEG with $r_s = 3.23$ (top row) and $r_s = 10$ (bottom row), both with $\theta = 1.0$. The finite $P$ error is shown for $\eta$-ensemble (left column), sign contribution (middle column) and total exchange-correlation contribution (right column). Uncorrected data (circles) for $N = 14$ are shown along with second-order polynomial fit (solid) and the $P \rightarrow \infty$ value (dotted). Corrected data points without (squares) and with (triangles) $\xi$-extrapolation show only small variations from constants, as indicated by their mean values (dashed). The $N = 30$ and $N = 66$ data in columns 1 and 2 have been shifted vertically, and FSC has been applied to $f_{xc}$.}
    \label{fig:FPC}
\end{figure*}

The computational cost of PIMC scales linearly with $P$, and to more efficiently model large $N$ it is desirable to keep $P$ as low as possible. However, a finite $P$ results in a systematic error~\cite{sakkos_JCP_2009,Dornheim_NJP_2015}, and convergence must be established. The convergence of $\Delta f_{\eta,\text{YR}}^{(\text{B})}$, $\Delta f_{S,\text{Ew}}$, and $f_{xc}$ for our two test cases with $r_s = 3.23$ and $r_s = 10$, are shown in Figure \ref{fig:FPC} for $N = 14$. Simulations were performed up to $P = 200$ using the primitive factorisation, and to achieve a systematic error below $0.1\%$ a $P \geq 20$ was required. No substantial dependence on $P$ was observed on $\Delta f_{a,\text{art-Ew}}^{(\text{B})}$.

A very similar systematic trend with $P$ was observed when the above study was carried out for $N = 30$ and $N = 66$, where finite-size errors were seen mainly to shift the result. Therefore, a second-order polynomial fit of the form  
\begin{equation}
    f(N, P) = p_0(N) + p_1 P^{-1} + p_2 P^{-2},
    \label{eq:FPC_fit_formula}
\end{equation}
where $p_0$, $p_1$ and $p_2$ are fitting coefficients, were carried out separately for $f = \Delta f_{\eta,\text{YR}}^{(\text{B})}$, $\Delta f_{S,\text{Ew}}$ and $f_{xc}$ for $N = 14$. By subtracting the $P$-dependence obtained from Equation \eqref{eq:FPC_fit_formula}, virtually all systematic errors are compensated for; see the demonstration in Figure \ref{fig:FPC}. This finite $P$ correction (FPC) method has been applied to all results for both $r_s = 3.23$ and $r_s = 10.0$.

\begin{table*}[t]
    \centering
    \caption{Coefficients obtained from the fits in Figure \ref{fig:FPC} using the functional form in Equation \eqref{eq:FPC_fit_formula}. The $p_0$-coefficients are omitted as they are $N$ dependent. Coefficients are given in units of the GDSMFB parametrisation~\cite{groth2017ab}. Results are shown for $r_s = 3.23$ and $r_s = 10$, both at $\theta = 1.0$ for the UEG.}
    \label{tab:FPC}
    \begin{tabular}{c||cc|cc}
         & \multicolumn{2}{c|}{$r_s = 3.23$} & \multicolumn{2}{c}{$r_s = 10.0$} \\
         & \multicolumn{1}{c}{$p_1$ [GDSMFB]} & \multicolumn{1}{c|}{$p_2$ [GDSMFB]} & \multicolumn{1}{c}{$p_1$ [GDSMFB]} & \multicolumn{1}{c}{$p_2$ [GDSMFB]} \\\hline
         $\Delta f_{\eta,\text{YR}}^{(\text{B})}$ & 0.021 & 0.30 & 0.0086 & 0.23 \\
         $\Delta f_{S,\text{Ew}}$ & -0.011 & -0.092 & -0.0023 & -0.030 \\\hline
         $f_{xc}$ & 0.0100 & 0.20 & 0.0060 & 0.20
    \end{tabular}
\end{table*}

The coefficients for the FPC are shown in Table \ref{tab:FPC}.  The quadratic correction dominates, unless $P \lesssim 10$. Furthermore, the finite $P$ error in the $\eta$-ensemble is typically larger than for the sign contribution. The coefficients for $f_{xc}$ are approximately the sum of the other two. However, a separate fit has been performed.

\section{Free energy for strongly coupled system: $r_s = 10$ \& $\theta = 1.0$}\label{sec:rs_10}

\begin{figure*}[t]
    \centering
    \includegraphics[width=\linewidth]{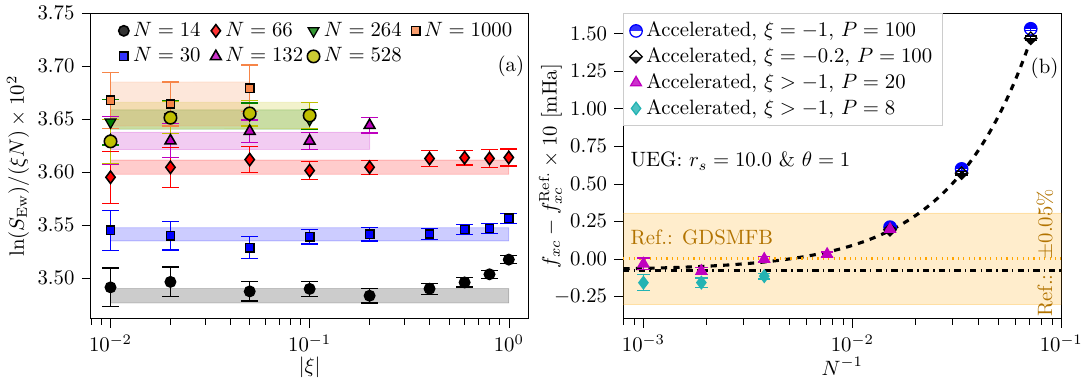}
    \caption{(a) Corresponding figure to Figure \mainrefsignextrapolation{} in main text for UEG at $r_s = 10$ and $\theta = 1.0$. The extrapolation technique is confirmed at a second condition. Some additional $\xi$-dependence on $a_S$ is seen for small $N$, although this is reduced for larger system sizes. (b) Corresponding figure to Figure \mainrefxcfreeenergy{} (main text) for UEG at $r_s = 10$ and $\theta = 1.0$. At these conditions the GDSMFB parametrisation gives $\Delta f_{xc} = -0.061120\,\text{Ha}$ (Ref.). Due to a reduced sign contribution, errors are further reduced compared the corresponding figure in the main text. On the scale of the  statistical error, the finite $P$ error can not fully corrected for by the procedure in Section \ref{sec:FPC} for $P = 8$ as compared to $P = 20$ at $N = 264$.}
    \label{fig:free_energy_rs_10}
\end{figure*}

The method presented is not restricted to the conditions discussed in the main text, and as a demonstration of this the corresponding computations for the UEG at $r_s = 10.0$ and $\theta = 1.0$ are shown in Figures \ref{fig:free_energy_rs_10}(a) and \ref{fig:free_energy_rs_10}(b). The structure of the sign extrapolation is generally the same as for $r_s = 3.23$, but a stronger $\xi$-dependence on $a_S$ is shown particularly for $N = 14$. However, this dependence is completely removed when a system size of $N = 66$ is reached, and above this point all $\xi$-dependence on $a_S$ can be neglected. This fits well to the heuristic explanation in terms of permutation cycles given in the main text.
Compared to the interaction contribution, the sign contribution decreases as $r_s$ increases and a stronger electron coupling is reached. Therefore, statistical errors -- primarily from the sign estimation -- are less prevalent and the estimates for the (finite-size corrected) free energy are well within $0.05\%$, see Figure \ref{fig:free_energy_rs_10}(b). The final estimation is $0.01\%$ lower than the GDSMFB parameterisation~\cite{groth2017ab}, which is very accurate within this regime.

Finite $P$ corrected results are shown for $P = 20$ and $P = 100$ in Figure \ref{fig:free_energy_rs_10}(b) with good agreement. However, for the $P = 8$ case, the correction formula in Equation \eqref{eq:FPC_fit_formula} is not able to fully correct the $P$-dependent error on the scale of $0.01\%$ and the data points disagree outside the estimate of the statistical error for $N = 264, 528, 1000$. Therefore, there is a lower bound on the needed $P$ to reach a desired accuracy. 

\twocolumn
\fi

\bibliography{ref}

\providecommand{\latin}[1]{#1}
\makeatletter
\providecommand{\doi}
  {\begingroup\let\do\@makeother\dospecials
  \catcode`\{=1 \catcode`\}=2 \doi@aux}
\providecommand{\doi@aux}[1]{\endgroup\texttt{#1}}
\makeatother
\providecommand*\mcitethebibliography{\thebibliography}
\csname @ifundefined\endcsname{endmcitethebibliography}  {\let\endmcitethebibliography\endthebibliography}{}
\begin{mcitethebibliography}{106}
\providecommand*\natexlab[1]{#1}
\providecommand*\mciteSetBstSublistMode[1]{}
\providecommand*\mciteSetBstMaxWidthForm[2]{}
\providecommand*\mciteBstWouldAddEndPuncttrue
  {\def\EndOfBibitem{\unskip.}}
\providecommand*\mciteBstWouldAddEndPunctfalse
  {\let\EndOfBibitem\relax}
\providecommand*\mciteSetBstMidEndSepPunct[3]{}
\providecommand*\mciteSetBstSublistLabelBeginEnd[3]{}
\providecommand*\EndOfBibitem{}
\mciteSetBstSublistMode{f}
\mciteSetBstMaxWidthForm{subitem}{(\alph{mcitesubitemcount})}
\mciteSetBstSublistLabelBeginEnd
  {\mcitemaxwidthsubitemform\space}
  {\relax}
  {\relax}

\bibitem[Giorgini \latin{et~al.}(2008)Giorgini, Pitaevskii, and Stringari]{giorgini2008theory}
Giorgini,~S.; Pitaevskii,~L.~P.; Stringari,~S. Theory of ultracold atomic Fermi gases. \emph{Reviews of Modern Physics} \textbf{2008}, \emph{80}, 1215\relax
\mciteBstWouldAddEndPuncttrue
\mciteSetBstMidEndSepPunct{\mcitedefaultmidpunct}
{\mcitedefaultendpunct}{\mcitedefaultseppunct}\relax
\EndOfBibitem
\bibitem[Ceperley(1992)]{ceperley1992path}
Ceperley,~D.~M. Path-integral calculations of normal liquid He 3. \emph{Physical Review Letters} \textbf{1992}, \emph{69}, 331\relax
\mciteBstWouldAddEndPuncttrue
\mciteSetBstMidEndSepPunct{\mcitedefaultmidpunct}
{\mcitedefaultendpunct}{\mcitedefaultseppunct}\relax
\EndOfBibitem
\bibitem[Reimann and Manninen(2002)Reimann, and Manninen]{reimann2002electronic}
Reimann,~S.~M.; Manninen,~M. Electronic structure of quantum dots. \emph{Reviews of Modern Physics} \textbf{2002}, \emph{74}, 1283\relax
\mciteBstWouldAddEndPuncttrue
\mciteSetBstMidEndSepPunct{\mcitedefaultmidpunct}
{\mcitedefaultendpunct}{\mcitedefaultseppunct}\relax
\EndOfBibitem
\bibitem[Dornheim and Yan(2022)Dornheim, and Yan]{dornheim2022abnormal}
Dornheim,~T.; Yan,~Y. Abnormal quantum moment of inertia and structural properties of electrons in 2D and 3D quantum dots: {An} \textit{ab initio} path-integral {Monte Carlo} study. \emph{New Journal of Physics} \textbf{2022}, \emph{24}, 113024\relax
\mciteBstWouldAddEndPuncttrue
\mciteSetBstMidEndSepPunct{\mcitedefaultmidpunct}
{\mcitedefaultendpunct}{\mcitedefaultseppunct}\relax
\EndOfBibitem
\bibitem[Lorenzen \latin{et~al.}(2014)Lorenzen, Becker, and Redmer]{lorenzen2014progress}
Lorenzen,~W.; Becker,~A.; Redmer,~R. In \emph{Frontiers and Challenges in Warm Dense Matter}; Graziani,~F., Desjarlais,~M.~P., Redmer,~R., Trickey,~S.~B., Eds.; Springer International Publishing: Cham, 2014; pp 203--234\relax
\mciteBstWouldAddEndPuncttrue
\mciteSetBstMidEndSepPunct{\mcitedefaultmidpunct}
{\mcitedefaultendpunct}{\mcitedefaultseppunct}\relax
\EndOfBibitem
\bibitem[Guillot(1999)]{guillot1999interiors}
Guillot,~T. Interiors of giant planets inside and outside the solar system. \emph{Science} \textbf{1999}, \emph{286}, 72--77\relax
\mciteBstWouldAddEndPuncttrue
\mciteSetBstMidEndSepPunct{\mcitedefaultmidpunct}
{\mcitedefaultendpunct}{\mcitedefaultseppunct}\relax
\EndOfBibitem
\bibitem[Fortney and Nettelmann(2010)Fortney, and Nettelmann]{fortney2010interior}
Fortney,~J.~J.; Nettelmann,~N. The Interior Structure, Composition, and Evolution of Giant Planets. \emph{Space Science Reviews} \textbf{2010}, \emph{152}, 423--447\relax
\mciteBstWouldAddEndPuncttrue
\mciteSetBstMidEndSepPunct{\mcitedefaultmidpunct}
{\mcitedefaultendpunct}{\mcitedefaultseppunct}\relax
\EndOfBibitem
\bibitem[Helled \latin{et~al.}(2020)Helled, Mazzola, and Redmer]{helled2020understanding}
Helled,~R.; Mazzola,~G.; Redmer,~R. Understanding dense hydrogen at planetary conditions. \emph{Nature Reviews Physics} \textbf{2020}, \emph{2}, 562--574\relax
\mciteBstWouldAddEndPuncttrue
\mciteSetBstMidEndSepPunct{\mcitedefaultmidpunct}
{\mcitedefaultendpunct}{\mcitedefaultseppunct}\relax
\EndOfBibitem
\bibitem[French \latin{et~al.}(2012)French, Becker, Lorenzen, Nettelmann, Bethkenhagen, Wicht, and Redmer]{french2012ab}
French,~M.; Becker,~A.; Lorenzen,~W.; Nettelmann,~N.; Bethkenhagen,~M.; Wicht,~J.; Redmer,~R. \textit{Ab initio} simulations for material properties along the {Jupiter} adiabat. \emph{The Astrophysical Journal Supplement Series} \textbf{2012}, \emph{202}, 5\relax
\mciteBstWouldAddEndPuncttrue
\mciteSetBstMidEndSepPunct{\mcitedefaultmidpunct}
{\mcitedefaultendpunct}{\mcitedefaultseppunct}\relax
\EndOfBibitem
\bibitem[Preising \latin{et~al.}(2023)Preising, French, Mankovich, Soubiran, and Redmer]{preising2023material}
Preising,~M.; French,~M.; Mankovich,~C.; Soubiran,~F.; Redmer,~R. Material properties of {Saturn’s} interior from \textit{ab initio} simulations. \emph{The Astrophysical Journal Supplement Series} \textbf{2023}, \emph{269}, 47\relax
\mciteBstWouldAddEndPuncttrue
\mciteSetBstMidEndSepPunct{\mcitedefaultmidpunct}
{\mcitedefaultendpunct}{\mcitedefaultseppunct}\relax
\EndOfBibitem
\bibitem[Kippenhahn \latin{et~al.}(2012)Kippenhahn, Weigert, and Weiss]{kippenhahn2012stellar}
Kippenhahn,~R.; Weigert,~A.; Weiss,~A. \emph{Stellar {S}tructure and {E}volution}, second ed. ed.; Springer: Heidelberg, 2012\relax
\mciteBstWouldAddEndPuncttrue
\mciteSetBstMidEndSepPunct{\mcitedefaultmidpunct}
{\mcitedefaultendpunct}{\mcitedefaultseppunct}\relax
\EndOfBibitem
\bibitem[Fortov(2009)]{fortov2009extreme}
Fortov,~V.~E. Extreme states of matter on Earth and in space. \emph{Physics-Uspekhi} \textbf{2009}, \emph{52}, 615\relax
\mciteBstWouldAddEndPuncttrue
\mciteSetBstMidEndSepPunct{\mcitedefaultmidpunct}
{\mcitedefaultendpunct}{\mcitedefaultseppunct}\relax
\EndOfBibitem
\bibitem[Chandrasekhar(1931)]{chandrasekhar1931density}
Chandrasekhar,~S. {XLVIII.} {The} density of white dwarf stars. \emph{The London, Edinburgh, and Dublin Philosophical Magazine and Journal of Science} \textbf{1931}, \emph{11}, 592--596\relax
\mciteBstWouldAddEndPuncttrue
\mciteSetBstMidEndSepPunct{\mcitedefaultmidpunct}
{\mcitedefaultendpunct}{\mcitedefaultseppunct}\relax
\EndOfBibitem
\bibitem[Chabrier \latin{et~al.}(2000)Chabrier, Brassard, Fontaine, and Saumon]{chabrier2000cooling}
Chabrier,~G.; Brassard,~P.; Fontaine,~G.; Saumon,~D. Cooling sequences and color-magnitude diagrams for cool white dwarfs with hydrogen atmospheres. \emph{The Astrophysical Journal} \textbf{2000}, \emph{543}, 216\relax
\mciteBstWouldAddEndPuncttrue
\mciteSetBstMidEndSepPunct{\mcitedefaultmidpunct}
{\mcitedefaultendpunct}{\mcitedefaultseppunct}\relax
\EndOfBibitem
\bibitem[Gudmundsson \latin{et~al.}(1983)Gudmundsson, Pethick, and Epstein]{gudmundsson1983structure}
Gudmundsson,~E.~H.; Pethick,~C.~J.; Epstein,~R.~I. Structure of neutron star envelopes. \emph{Astrophysical Journal} \textbf{1983}, \emph{272}, 286--300\relax
\mciteBstWouldAddEndPuncttrue
\mciteSetBstMidEndSepPunct{\mcitedefaultmidpunct}
{\mcitedefaultendpunct}{\mcitedefaultseppunct}\relax
\EndOfBibitem
\bibitem[Haensel \latin{et~al.}(2007)Haensel, Potekhin, and Yakovlev]{haensel2007neutron}
Haensel,~P.; Potekhin,~A.~Y.; Yakovlev,~D.~G. \emph{Neutron {S}tars 1: {Equation} of {S}tate and {S}tructure}; Springer: New York, NY, 2007\relax
\mciteBstWouldAddEndPuncttrue
\mciteSetBstMidEndSepPunct{\mcitedefaultmidpunct}
{\mcitedefaultendpunct}{\mcitedefaultseppunct}\relax
\EndOfBibitem
\bibitem[Nuckolls \latin{et~al.}(1972)Nuckolls, Wood, Thiessen, and Zimmerman]{nuckolls1972laser}
Nuckolls,~J.; Wood,~L.; Thiessen,~A.; Zimmerman,~G. Laser compression of matter to super-high densities: {Thermonuclear} ({CTR}) applications. \emph{Nature} \textbf{1972}, \emph{239}, 139--142\relax
\mciteBstWouldAddEndPuncttrue
\mciteSetBstMidEndSepPunct{\mcitedefaultmidpunct}
{\mcitedefaultendpunct}{\mcitedefaultseppunct}\relax
\EndOfBibitem
\bibitem[Betti and Hurricane(2016)Betti, and Hurricane]{betti2016inertial}
Betti,~R.; Hurricane,~O.~A. Inertial-confinement fusion with lasers. \emph{Nature Physics} \textbf{2016}, \emph{12}, 435--448\relax
\mciteBstWouldAddEndPuncttrue
\mciteSetBstMidEndSepPunct{\mcitedefaultmidpunct}
{\mcitedefaultendpunct}{\mcitedefaultseppunct}\relax
\EndOfBibitem
\bibitem[Hurricane \latin{et~al.}(2023)Hurricane, Patel, Betti, Froula, Regan, Slutz, Gomez, and Sweeney]{hurricane2023physics}
Hurricane,~O.~A.; Patel,~P.~K.; Betti,~R.; Froula,~D.~H.; Regan,~S.~P.; Slutz,~S.~A.; Gomez,~M.~R.; Sweeney,~M.~A. Physics principles of inertial confinement fusion and {US} program overview. \emph{Reviews of Modern Physics} \textbf{2023}, \emph{95}, 025005\relax
\mciteBstWouldAddEndPuncttrue
\mciteSetBstMidEndSepPunct{\mcitedefaultmidpunct}
{\mcitedefaultendpunct}{\mcitedefaultseppunct}\relax
\EndOfBibitem
\bibitem[Miao \latin{et~al.}(2020)Miao, Sun, Zurek, and Lin]{miao2020chemistry}
Miao,~M.; Sun,~Y.; Zurek,~E.; Lin,~H. Chemistry under high pressure. \emph{Nature Reviews Chemistry} \textbf{2020}, \emph{4}, 508--527\relax
\mciteBstWouldAddEndPuncttrue
\mciteSetBstMidEndSepPunct{\mcitedefaultmidpunct}
{\mcitedefaultendpunct}{\mcitedefaultseppunct}\relax
\EndOfBibitem
\bibitem[Abu-Shawareb \latin{et~al.}(2022)Abu-Shawareb, Acree, Adams, Adams, Addis, Aden, Adrian, Afeyan, Aggleton, Aghaian, \latin{et~al.} others]{abu2022lawson}
Abu-Shawareb,~H.; Acree,~R.; Adams,~P.; Adams,~J.; Addis,~B.; Aden,~R.; Adrian,~P.; Afeyan,~B.~B.; Aggleton,~M.; Aghaian,~L. \latin{et~al.}  Lawson criterion for ignition exceeded in an inertial fusion experiment. \emph{Physical Review Letters} \textbf{2022}, \emph{129}, 075001\relax
\mciteBstWouldAddEndPuncttrue
\mciteSetBstMidEndSepPunct{\mcitedefaultmidpunct}
{\mcitedefaultendpunct}{\mcitedefaultseppunct}\relax
\EndOfBibitem
\bibitem[Bonitz \latin{et~al.}(2020)Bonitz, Dornheim, Moldabekov, Zhang, Hamann, K{\"a}hlert, Filinov, Ramakrishna, and Vorberger]{bonitz2020ab}
Bonitz,~M.; Dornheim,~T.; Moldabekov,~Z.~A.; Zhang,~S.; Hamann,~P.; K{\"a}hlert,~H.; Filinov,~A.; Ramakrishna,~K.; Vorberger,~J. \textit{Ab initio} simulation of warm dense matter. \emph{Physics of Plasmas} \textbf{2020}, \emph{27}, 042710\relax
\mciteBstWouldAddEndPuncttrue
\mciteSetBstMidEndSepPunct{\mcitedefaultmidpunct}
{\mcitedefaultendpunct}{\mcitedefaultseppunct}\relax
\EndOfBibitem
\bibitem[Vorberger \latin{et~al.}(2025)Vorberger, Graziani, Riley, Baczewski, Baraffe, Bethkenhagen, Blouin, Böhme, Bonitz, Bussmann, Casner, Cayzac, Celliers, Chabrier, Chamel, Chapman, Chen, Clérouin, Collins, Coppari, Döppner, Dornheim, Fletcher, Gericke, Glenzer, Goncharov, Gregori, Hamel, Hansen, Hartley, Hu, Hurricane, Karasiev, Kas, Kettle, Kluge, Knudson, Kononov, á, Kraus, Kritcher, Malko, Massacrier, Militzer, Moldabekov, Murillo, Nagler, Nettelmann, Neumayer, Ofori-Okai, Oleynik, Preising, Pribram-Jones, Ramazanov, Ravasio, Redmer, Rethfeld, Robinson, Röpke, Soubiran, Starrett, Steinle-Neumann, Sterne, Tanaka, Thompson, Trickey, Vinci, Vinko, Wang, White, White, Zastrau, Zurek, and Tolias]{vorberger2025roadmapwarmdensematter}
Vorberger,~J.; Graziani,~F.; Riley,~D.; Baczewski,~A.~D.; Baraffe,~I.; Bethkenhagen,~M.; Blouin,~S.; Böhme,~M.~P.; Bonitz,~M.; Bussmann,~M. \latin{et~al.}  Roadmap for warm dense matter physics. 2025; \url{https://arxiv.org/abs/2505.02494}\relax
\mciteBstWouldAddEndPuncttrue
\mciteSetBstMidEndSepPunct{\mcitedefaultmidpunct}
{\mcitedefaultendpunct}{\mcitedefaultseppunct}\relax
\EndOfBibitem
\bibitem[Wahl \latin{et~al.}(2017)Wahl, Hubbard, Militzer, Guillot, Miguel, Movshovitz, Kaspi, Helled, Reese, Galanti, \latin{et~al.} others]{wahl2017comparing}
Wahl,~S.~M.; Hubbard,~W.~B.; Militzer,~B.; Guillot,~T.; Miguel,~Y.; Movshovitz,~N.; Kaspi,~Y.; Helled,~R.; Reese,~D.; Galanti,~E. \latin{et~al.}  Comparing {Jupiter} interior structure models to {Juno} gravity measurements and the role of a dilute core. \emph{Geophysical Research Letters} \textbf{2017}, \emph{44}, 4649--4659\relax
\mciteBstWouldAddEndPuncttrue
\mciteSetBstMidEndSepPunct{\mcitedefaultmidpunct}
{\mcitedefaultendpunct}{\mcitedefaultseppunct}\relax
\EndOfBibitem
\bibitem[Hu \latin{et~al.}(2015)Hu, Goncharov, Boehly, McCrory, Skupsky, Collins, Kress, and Militzer]{hu2015impact}
Hu,~S.~X.; Goncharov,~V.~N.; Boehly,~T.~R.; McCrory,~R.~L.; Skupsky,~S.; Collins,~L.~A.; Kress,~J.~D.; Militzer,~B. Impact of first-principles properties of deuterium--tritium on inertial confinement fusion target designs. \emph{Physics of Plasmas} \textbf{2015}, \emph{22}, 056304\relax
\mciteBstWouldAddEndPuncttrue
\mciteSetBstMidEndSepPunct{\mcitedefaultmidpunct}
{\mcitedefaultendpunct}{\mcitedefaultseppunct}\relax
\EndOfBibitem
\bibitem[Bonitz \latin{et~al.}(2024)Bonitz, Vorberger, Bethkenhagen, Böhme, Ceperley, Filinov, Gawne, Graziani, Gregori, Hamann, Hansen, Holzmann, Hu, Kählert, Karasiev, Kleinschmidt, Kordts, Makait, Militzer, Moldabekov, Pierleoni, Preising, Ramakrishna, Redmer, Schwalbe, Svensson, and Dornheim]{bonitz2024principles}
Bonitz,~M.; Vorberger,~J.; Bethkenhagen,~M.; Böhme,~M.; Ceperley,~D.; Filinov,~A.; Gawne,~T.; Graziani,~F.; Gregori,~G.; Hamann,~P. \latin{et~al.}  Toward first principles-based simulations of dense hydrogen. \emph{Physics of Plasmas} \textbf{2024}, \emph{31}, 110501\relax
\mciteBstWouldAddEndPuncttrue
\mciteSetBstMidEndSepPunct{\mcitedefaultmidpunct}
{\mcitedefaultendpunct}{\mcitedefaultseppunct}\relax
\EndOfBibitem
\bibitem[Hohenberg and Kohn(1964)Hohenberg, and Kohn]{hohenberg1964inhomogeneous}
Hohenberg,~P.; Kohn,~W. Inhomogeneous electron gas. \emph{Physical Review} \textbf{1964}, \emph{136}, B864\relax
\mciteBstWouldAddEndPuncttrue
\mciteSetBstMidEndSepPunct{\mcitedefaultmidpunct}
{\mcitedefaultendpunct}{\mcitedefaultseppunct}\relax
\EndOfBibitem
\bibitem[Mermin(1965)]{mermin1965thermal}
Mermin,~N.~D. Thermal properties of the inhomogeneous electron gas. \emph{Physical Review} \textbf{1965}, \emph{137}, A1441\relax
\mciteBstWouldAddEndPuncttrue
\mciteSetBstMidEndSepPunct{\mcitedefaultmidpunct}
{\mcitedefaultendpunct}{\mcitedefaultseppunct}\relax
\EndOfBibitem
\bibitem[Rapaport(2004)]{rapaport2004art}
Rapaport,~D.~C. \emph{The {A}rt of {M}olecular {D}ynamics {S}imulation}; Cambridge University Press: Cambridge, 2004\relax
\mciteBstWouldAddEndPuncttrue
\mciteSetBstMidEndSepPunct{\mcitedefaultmidpunct}
{\mcitedefaultendpunct}{\mcitedefaultseppunct}\relax
\EndOfBibitem
\bibitem[Perdew and Zunger(1981)Perdew, and Zunger]{perdew1981self}
Perdew,~J.~P.; Zunger,~A. Self-interaction correction to density-functional approximations for many-electron systems. \emph{Physical Review B} \textbf{1981}, \emph{23}, 5048\relax
\mciteBstWouldAddEndPuncttrue
\mciteSetBstMidEndSepPunct{\mcitedefaultmidpunct}
{\mcitedefaultendpunct}{\mcitedefaultseppunct}\relax
\EndOfBibitem
\bibitem[Karasiev \latin{et~al.}(2014)Karasiev, Sjostrom, Dufty, and Trickey]{karasiev2014accurate}
Karasiev,~V.~V.; Sjostrom,~T.; Dufty,~J.; Trickey,~S. Accurate homogeneous electron gas exchange-correlation free energy for local spin-density calculations. \emph{Physical Review Letters} \textbf{2014}, \emph{112}, 076403\relax
\mciteBstWouldAddEndPuncttrue
\mciteSetBstMidEndSepPunct{\mcitedefaultmidpunct}
{\mcitedefaultendpunct}{\mcitedefaultseppunct}\relax
\EndOfBibitem
\bibitem[Dornheim \latin{et~al.}(2018)Dornheim, Groth, and Bonitz]{dornheim2018uniform}
Dornheim,~T.; Groth,~S.; Bonitz,~M. The uniform electron gas at warm dense matter conditions. \emph{Physics Reports} \textbf{2018}, \emph{744}, 1--86\relax
\mciteBstWouldAddEndPuncttrue
\mciteSetBstMidEndSepPunct{\mcitedefaultmidpunct}
{\mcitedefaultendpunct}{\mcitedefaultseppunct}\relax
\EndOfBibitem
\bibitem[Ceperley(1995)]{ceperley1995path}
Ceperley,~D.~M. Path integrals in the theory of condensed helium. \emph{Reviews of Modern Physics} \textbf{1995}, \emph{67}, 279\relax
\mciteBstWouldAddEndPuncttrue
\mciteSetBstMidEndSepPunct{\mcitedefaultmidpunct}
{\mcitedefaultendpunct}{\mcitedefaultseppunct}\relax
\EndOfBibitem
\bibitem[B\"ohme \latin{et~al.}(2022)B\"ohme, Moldabekov, Vorberger, and Dornheim]{Bohme_PRL_2022}
B\"ohme,~M.; Moldabekov,~Z.~A.; Vorberger,~J.; Dornheim,~T. Static electronic density response of warm dense hydrogen: \textit{Ab initio} path integral Monte Carlo Simulations. \emph{Phys. Rev. Lett.} \textbf{2022}, \emph{129}, 066402\relax
\mciteBstWouldAddEndPuncttrue
\mciteSetBstMidEndSepPunct{\mcitedefaultmidpunct}
{\mcitedefaultendpunct}{\mcitedefaultseppunct}\relax
\EndOfBibitem
\bibitem[Dornheim(2019)]{dornheim2019fermion}
Dornheim,~T. Fermion sign problem in path integral {M}onte {C}arlo simulations: {Quantum} dots, ultracold atoms, and warm dense matter. \emph{Physical Review E} \textbf{2019}, \emph{100}, 023307\relax
\mciteBstWouldAddEndPuncttrue
\mciteSetBstMidEndSepPunct{\mcitedefaultmidpunct}
{\mcitedefaultendpunct}{\mcitedefaultseppunct}\relax
\EndOfBibitem
\bibitem[Troyer and Wiese(2005)Troyer, and Wiese]{troyer}
Troyer,~M.; Wiese,~U.~J. Computational Complexity and Fundamental Limitations to Fermionic Quantum {M}onte {C}arlo Simulations. \emph{Physical Review Letters} \textbf{2005}, \emph{94}, 170201\relax
\mciteBstWouldAddEndPuncttrue
\mciteSetBstMidEndSepPunct{\mcitedefaultmidpunct}
{\mcitedefaultendpunct}{\mcitedefaultseppunct}\relax
\EndOfBibitem
\bibitem[Hatano(1994)]{hatano1994data}
Hatano,~N. Data analysis for quantum {Monte Carlo} simulations with the negative-sign problem. \emph{Journal of the Physical Society of Japan} \textbf{1994}, \emph{63}, 1691--1697\relax
\mciteBstWouldAddEndPuncttrue
\mciteSetBstMidEndSepPunct{\mcitedefaultmidpunct}
{\mcitedefaultendpunct}{\mcitedefaultseppunct}\relax
\EndOfBibitem
\bibitem[Xiong and Xiong(2022)Xiong, and Xiong]{xiong2022thermodynamic}
Xiong,~Y.; Xiong,~H. On the thermodynamic properties of fictitious identical particles and the application to fermion sign problem. \emph{The Journal of Chemical Physics} \textbf{2022}, \emph{157}, 094112\relax
\mciteBstWouldAddEndPuncttrue
\mciteSetBstMidEndSepPunct{\mcitedefaultmidpunct}
{\mcitedefaultendpunct}{\mcitedefaultseppunct}\relax
\EndOfBibitem
\bibitem[Dornheim \latin{et~al.}(2023)Dornheim, Tolias, Groth, Moldabekov, Vorberger, and Hirshberg]{dornheim2023fermionic}
Dornheim,~T.; Tolias,~P.; Groth,~S.; Moldabekov,~Z.~A.; Vorberger,~J.; Hirshberg,~B. Fermionic physics from \textit{ab initio} path integral {Monte Carlo} simulations of fictitious identical particles. \emph{The Journal of Chemical Physics} \textbf{2023}, \emph{159}, 164113\relax
\mciteBstWouldAddEndPuncttrue
\mciteSetBstMidEndSepPunct{\mcitedefaultmidpunct}
{\mcitedefaultendpunct}{\mcitedefaultseppunct}\relax
\EndOfBibitem
\bibitem[Dornheim \latin{et~al.}(2024)Dornheim, Schwalbe, Moldabekov, Vorberger, and Tolias]{Dornheim_JPCL_2024}
Dornheim,~T.; Schwalbe,~S.; Moldabekov,~Z.~A.; Vorberger,~J.; Tolias,~P. \textit{Ab initio} path integral {Monte Carlo} simulations of the uniform electron gas on large length scales. \emph{The Journal of Physical Chemistry Letters} \textbf{2024}, \emph{15}, 1305--1313\relax
\mciteBstWouldAddEndPuncttrue
\mciteSetBstMidEndSepPunct{\mcitedefaultmidpunct}
{\mcitedefaultendpunct}{\mcitedefaultseppunct}\relax
\EndOfBibitem
\bibitem[Xiong and Xiong(2023)Xiong, and Xiong]{xiong2023thermodynamics}
Xiong,~Y.; Xiong,~H. Thermodynamics of fermions at any temperature based on parametrized partition function. \emph{Physical Review E} \textbf{2023}, \emph{107}, 055308\relax
\mciteBstWouldAddEndPuncttrue
\mciteSetBstMidEndSepPunct{\mcitedefaultmidpunct}
{\mcitedefaultendpunct}{\mcitedefaultseppunct}\relax
\EndOfBibitem
\bibitem[Morresi and Garberoglio(2025)Morresi, and Garberoglio]{morresi2025normal}
Morresi,~T.; Garberoglio,~G. Normal liquid $^{3}$He studied by path-integral {Monte Carlo} with a parametrized partition function. \emph{Physical Review B} \textbf{2025}, \emph{111}, 014521\relax
\mciteBstWouldAddEndPuncttrue
\mciteSetBstMidEndSepPunct{\mcitedefaultmidpunct}
{\mcitedefaultendpunct}{\mcitedefaultseppunct}\relax
\EndOfBibitem
\bibitem[Dornheim \latin{et~al.}(2024)Dornheim, Schwalbe, Moldabekov, Vorberger, and Tolias]{dornheim2024ab}
Dornheim,~T.; Schwalbe,~S.; Moldabekov,~Z.~A.; Vorberger,~J.; Tolias,~P. \textit{Ab initio} path integral {Monte Carlo} simulations of the uniform electron gas on large length scales. \emph{The Journal of Physical Chemistry Letters} \textbf{2024}, \emph{15}, 1305--1313\relax
\mciteBstWouldAddEndPuncttrue
\mciteSetBstMidEndSepPunct{\mcitedefaultmidpunct}
{\mcitedefaultendpunct}{\mcitedefaultseppunct}\relax
\EndOfBibitem
\bibitem[Dornheim \latin{et~al.}(2024)Dornheim, Schwalbe, B{\"o}hme, Moldabekov, Vorberger, and Tolias]{dornheim2024ab_b}
Dornheim,~T.; Schwalbe,~S.; B{\"o}hme,~M.~P.; Moldabekov,~Z.~A.; Vorberger,~J.; Tolias,~P. \textit{Ab initio} path integral {Monte Carlo} simulations of warm dense two-component systems without fixed nodes: {Structural} properties. \emph{The Journal of Chemical Physics} \textbf{2024}, \emph{160}, 164111\relax
\mciteBstWouldAddEndPuncttrue
\mciteSetBstMidEndSepPunct{\mcitedefaultmidpunct}
{\mcitedefaultendpunct}{\mcitedefaultseppunct}\relax
\EndOfBibitem
\bibitem[Dornheim \latin{et~al.}(2025)Dornheim, D{\"o}ppner, Tolias, B{\"o}hme, Fletcher, Gawne, Graziani, Kraus, MacDonald, Moldabekov, \latin{et~al.} others]{dornheim2025unraveling}
Dornheim,~T.; D{\"o}ppner,~T.; Tolias,~P.; B{\"o}hme,~M.~P.; Fletcher,~L.~B.; Gawne,~T.; Graziani,~F.~R.; Kraus,~D.; MacDonald,~M.~J.; Moldabekov,~Z.~A. \latin{et~al.}  Unraveling electronic correlations in warm dense quantum plasmas. \emph{Nature Communications} \textbf{2025}, \emph{16}, 5103\relax
\mciteBstWouldAddEndPuncttrue
\mciteSetBstMidEndSepPunct{\mcitedefaultmidpunct}
{\mcitedefaultendpunct}{\mcitedefaultseppunct}\relax
\EndOfBibitem
\bibitem[Dornheim \latin{et~al.}(2024)Dornheim, Schwalbe, Tolias, B{\"o}hme, Moldabekov, and Vorberger]{dornheim2024ab_c}
Dornheim,~T.; Schwalbe,~S.; Tolias,~P.; B{\"o}hme,~M.~P.; Moldabekov,~Z.~A.; Vorberger,~J. \textit{Ab initio} density response and local field factor of warm dense hydrogen. \emph{Matter and Radiation at Extremes} \textbf{2024}, \emph{9}, 057401\relax
\mciteBstWouldAddEndPuncttrue
\mciteSetBstMidEndSepPunct{\mcitedefaultmidpunct}
{\mcitedefaultendpunct}{\mcitedefaultseppunct}\relax
\EndOfBibitem
\bibitem[Dornheim \latin{et~al.}(2025)Dornheim, Moldabekov, Schwalbe, Tolias, and Vorberger]{dornheim2025fermionic}
Dornheim,~T.; Moldabekov,~Z.; Schwalbe,~S.; Tolias,~P.; Vorberger,~J. Fermionic free energies from \textit{ab initio} path integral {Monte Carlo} simulations of fictitious identical particles. \emph{arXiv preprint arXiv:2502.15288} \textbf{2025}, \relax
\mciteBstWouldAddEndPunctfalse
\mciteSetBstMidEndSepPunct{\mcitedefaultmidpunct}
{}{\mcitedefaultseppunct}\relax
\EndOfBibitem
\bibitem[Gupta and Rajagopal(1982)Gupta, and Rajagopal]{gupta1982density}
Gupta,~U.; Rajagopal,~A.~K. Density functional formalism at finite temperatures with some applications. \emph{Physics Reports} \textbf{1982}, \emph{87}, 259--311\relax
\mciteBstWouldAddEndPuncttrue
\mciteSetBstMidEndSepPunct{\mcitedefaultmidpunct}
{\mcitedefaultendpunct}{\mcitedefaultseppunct}\relax
\EndOfBibitem
\bibitem[Smith \latin{et~al.}(2018)Smith, Sagredo, and Burke]{smith2018warming}
Smith,~J.~C.; Sagredo,~F.; Burke,~K. In \emph{Frontiers of Quantum Chemistry}; W{\'o}jcik,~M.~J., Nakatsuji,~H., Kirtman,~B., Ozaki,~Y., Eds.; Springer Singapore: Singapore, 2018; pp 249--271\relax
\mciteBstWouldAddEndPuncttrue
\mciteSetBstMidEndSepPunct{\mcitedefaultmidpunct}
{\mcitedefaultendpunct}{\mcitedefaultseppunct}\relax
\EndOfBibitem
\bibitem[Sjostrom and Daligault(2014)Sjostrom, and Daligault]{sjostrom2014gradient}
Sjostrom,~T.; Daligault,~J. Gradient corrections to the exchange-correlation free energy. \emph{Physical Review B} \textbf{2014}, \emph{90}, 155109\relax
\mciteBstWouldAddEndPuncttrue
\mciteSetBstMidEndSepPunct{\mcitedefaultmidpunct}
{\mcitedefaultendpunct}{\mcitedefaultseppunct}\relax
\EndOfBibitem
\bibitem[Karasiev \latin{et~al.}(2016)Karasiev, Calder\'{\i}n, and Trickey]{karasiev2016importance}
Karasiev,~V.~V.; Calder\'{\i}n,~L.; Trickey,~S.~B. Importance of finite-temperature exchange correlation for warm dense matter calculations. \emph{Physical Review E} \textbf{2016}, \emph{93}, 063207\relax
\mciteBstWouldAddEndPuncttrue
\mciteSetBstMidEndSepPunct{\mcitedefaultmidpunct}
{\mcitedefaultendpunct}{\mcitedefaultseppunct}\relax
\EndOfBibitem
\bibitem[Ramakrishna \latin{et~al.}(2020)Ramakrishna, Dornheim, and Vorberger]{ramakrishna2020influence}
Ramakrishna,~K.; Dornheim,~T.; Vorberger,~J. Influence of finite temperature exchange-correlation effects in hydrogen. \emph{Physical Review B} \textbf{2020}, \emph{101}, 195129\relax
\mciteBstWouldAddEndPuncttrue
\mciteSetBstMidEndSepPunct{\mcitedefaultmidpunct}
{\mcitedefaultendpunct}{\mcitedefaultseppunct}\relax
\EndOfBibitem
\bibitem[Alfe \latin{et~al.}(1999)Alfe, Gillan, and Price]{alfe1999melting}
Alfe,~D.; Gillan,~M.~J.; Price,~G.~D. The melting curve of iron at the pressures of the {Earth's} core from \textit{ab initio} calculations. \emph{Nature} \textbf{1999}, \emph{401}, 462--464\relax
\mciteBstWouldAddEndPuncttrue
\mciteSetBstMidEndSepPunct{\mcitedefaultmidpunct}
{\mcitedefaultendpunct}{\mcitedefaultseppunct}\relax
\EndOfBibitem
\bibitem[Wilson and Militzer(2011)Wilson, and Militzer]{wilson2011solubility}
Wilson,~H.~F.; Militzer,~B. Solubility of water ice in metallic hydrogen: {Consequences} for core erosion in gas giant planets. \emph{The Astrophysical Journal} \textbf{2011}, \emph{745}, 54\relax
\mciteBstWouldAddEndPuncttrue
\mciteSetBstMidEndSepPunct{\mcitedefaultmidpunct}
{\mcitedefaultendpunct}{\mcitedefaultseppunct}\relax
\EndOfBibitem
\bibitem[Wu \latin{et~al.}(2021)Wu, Gonz{\'a}lez-Cataldo, and Militzer]{wu2021high}
Wu,~J.; Gonz{\'a}lez-Cataldo,~F.; Militzer,~B. High-pressure phase diagram of beryllium from \textit{ab initio} free-energy calculations. \emph{Physical Review B} \textbf{2021}, \emph{104}, 014103\relax
\mciteBstWouldAddEndPuncttrue
\mciteSetBstMidEndSepPunct{\mcitedefaultmidpunct}
{\mcitedefaultendpunct}{\mcitedefaultseppunct}\relax
\EndOfBibitem
\bibitem[Militzer \latin{et~al.}(2021)Militzer, Gonz{\'a}lez-Cataldo, Zhang, Driver, and Soubiran]{militzer2021first}
Militzer,~B.; Gonz{\'a}lez-Cataldo,~F.; Zhang,~S.; Driver,~K.~P.; Soubiran,~F. First-principles equation of state database for warm dense matter computation. \emph{Physical Review E} \textbf{2021}, \emph{103}, 013203\relax
\mciteBstWouldAddEndPuncttrue
\mciteSetBstMidEndSepPunct{\mcitedefaultmidpunct}
{\mcitedefaultendpunct}{\mcitedefaultseppunct}\relax
\EndOfBibitem
\bibitem[More \latin{et~al.}(1988)More, Warren, Young, and Zimmerman]{more1988new}
More,~R.~M.; Warren,~K.; Young,~D.; Zimmerman,~G. A new quotidian equation of state (QEOS) for hot dense matter. \emph{The Physics of fluids} \textbf{1988}, \emph{31}, 3059--3078\relax
\mciteBstWouldAddEndPuncttrue
\mciteSetBstMidEndSepPunct{\mcitedefaultmidpunct}
{\mcitedefaultendpunct}{\mcitedefaultseppunct}\relax
\EndOfBibitem
\bibitem[Lyon and Johnson(1995)Lyon, and Johnson]{lyon1995sesame}
Lyon,~S.~P.; Johnson,~J.~D. \emph{Technical Report {No. LA-UR-92-3407}}; Los Alamos National Laboratory: Los Alamos, 1995\relax
\mciteBstWouldAddEndPuncttrue
\mciteSetBstMidEndSepPunct{\mcitedefaultmidpunct}
{\mcitedefaultendpunct}{\mcitedefaultseppunct}\relax
\EndOfBibitem
\bibitem[Pople(1999)]{pople1999nobel}
Pople,~J.~A. Nobel lecture: {Quantum} chemical models. \emph{Reviews of Modern Physics} \textbf{1999}, \emph{71}, 1267\relax
\mciteBstWouldAddEndPuncttrue
\mciteSetBstMidEndSepPunct{\mcitedefaultmidpunct}
{\mcitedefaultendpunct}{\mcitedefaultseppunct}\relax
\EndOfBibitem
\bibitem[Zastrau \latin{et~al.}(2021)Zastrau, Appel, Baehtz, Baehr, Batchelor, Bergh{\"{a}}user, Banjafar, Brambrink, Cerantola, Cowan, Damker, Dietrich, {Di Dio Cafiso}, Dreyer, Engel, Feldmann, Findeisen, Foese, Fulla-Marsa, G{\"{o}}de, Hassan, Hauser, Herrmannsd{\"{o}}rfer, H{\"{o}}ppner, Kaa, Kaever, Kn{\"{o}}fel, Kon{\^{o}}pkov{\'{a}}, {Laso Garc{\'{i}}a}, Liermann, Mainberger, Makita, Martens, McBride, M{\"{o}}ller, Nakatsutsumi, Pelka, Plueckthun, Prescher, Preston, R{\"{o}}per, Schmidt, Seidel, Schwinkendorf, Schoelmerich, Schramm, Schropp, Strohm, Sukharnikov, Talkovski, Thorpe, Toncian, Toncian, Wollenweber, Yamamoto, and Tschentscher]{Zastrau2021}
Zastrau,~U.; Appel,~K.; Baehtz,~C.; Baehr,~O.; Batchelor,~L.; Bergh{\"{a}}user,~A.; Banjafar,~M.; Brambrink,~E.; Cerantola,~V.; Cowan,~T.~E. \latin{et~al.}  {The High Energy Density Scientific Instrument at the European XFEL}. \emph{Journal of Synchrotron Radiation} \textbf{2021}, \emph{28}, 1393--1416\relax
\mciteBstWouldAddEndPuncttrue
\mciteSetBstMidEndSepPunct{\mcitedefaultmidpunct}
{\mcitedefaultendpunct}{\mcitedefaultseppunct}\relax
\EndOfBibitem
\bibitem[Fletcher \latin{et~al.}(2022)Fletcher, Vorberger, Schumaker, Ruyer, Goede, Galtier, Zastrau, Alves, Baalrud, Baggott, Barbrel, Chen, Döppner, Gauthier, Granados, Kim, Kraus, Lee, MacDonald, Mishra, Pelka, Ravasio, Roedel, Fry, Redmer, Fiuza, Gericke, and Glenzer]{Fletcher_Frontiers_2022}
Fletcher,~L.~B.; Vorberger,~J.; Schumaker,~W.; Ruyer,~C.; Goede,~S.; Galtier,~E.; Zastrau,~U.; Alves,~E.~P.; Baalrud,~S.~D.; Baggott,~R.~A. \latin{et~al.}  Electron-Ion Temperature Relaxation in Warm Dense Hydrogen Observed With Picosecond Resolved X-Ray Scattering. \emph{Frontiers in Physics} \textbf{2022}, \emph{10}\relax
\mciteBstWouldAddEndPuncttrue
\mciteSetBstMidEndSepPunct{\mcitedefaultmidpunct}
{\mcitedefaultendpunct}{\mcitedefaultseppunct}\relax
\EndOfBibitem
\bibitem[Hamann \latin{et~al.}(2023)Hamann, Kordts, Filinov, Bonitz, Dornheim, and Vorberger]{Hamann_PRR_2023}
Hamann,~P.; Kordts,~L.; Filinov,~A.; Bonitz,~M.; Dornheim,~T.; Vorberger,~J. Prediction of a roton-type feature in warm dense hydrogen. \emph{Phys. Rev. Res.} \textbf{2023}, \emph{5}, 033039\relax
\mciteBstWouldAddEndPuncttrue
\mciteSetBstMidEndSepPunct{\mcitedefaultmidpunct}
{\mcitedefaultendpunct}{\mcitedefaultseppunct}\relax
\EndOfBibitem
\bibitem[Frenkel and Smit(2002)Frenkel, and Smit]{frenkel2002understanding}
Frenkel,~D.; Smit,~B. \emph{Understanding Molecular Simulation: From Algorithms to Applications}, 2nd ed.; Academic Press: San Diego, USA, 2002; Chapter Free Energy Calculations, pp 167--200\relax
\mciteBstWouldAddEndPuncttrue
\mciteSetBstMidEndSepPunct{\mcitedefaultmidpunct}
{\mcitedefaultendpunct}{\mcitedefaultseppunct}\relax
\EndOfBibitem
\bibitem[Zwanzig(1954)]{zwanzig1954high}
Zwanzig,~R.~W. High-temperature equation of state by a perturbation method. {I.} {Nonpolar} gases. \emph{The Journal of Chemical Physics} \textbf{1954}, \emph{22}, 1420--1426\relax
\mciteBstWouldAddEndPuncttrue
\mciteSetBstMidEndSepPunct{\mcitedefaultmidpunct}
{\mcitedefaultendpunct}{\mcitedefaultseppunct}\relax
\EndOfBibitem
\bibitem[Pribram-Jones \latin{et~al.}(2016)Pribram-Jones, Grabowski, and Burke]{pribram2016thermal}
Pribram-Jones,~A.; Grabowski,~P.~E.; Burke,~K. Thermal density functional theory: {Time}-dependent linear response and approximate functionals from the fluctuation-dissipation theorem. \emph{Physical Review Letters} \textbf{2016}, \emph{116}, 233001\relax
\mciteBstWouldAddEndPuncttrue
\mciteSetBstMidEndSepPunct{\mcitedefaultmidpunct}
{\mcitedefaultendpunct}{\mcitedefaultseppunct}\relax
\EndOfBibitem
\bibitem[Dornheim \latin{et~al.}(2025)Dornheim, Moldabekov, Schwalbe, and Vorberger]{dornheim2025direct}
Dornheim,~T.; Moldabekov,~Z.~A.; Schwalbe,~S.; Vorberger,~J. Direct free energy calculation from \textit{ab initio} path integral {Monte Carlo} simulations of warm dense matter. \emph{Physical Review B} \textbf{2025}, \emph{111}, L041114\relax
\mciteBstWouldAddEndPuncttrue
\mciteSetBstMidEndSepPunct{\mcitedefaultmidpunct}
{\mcitedefaultendpunct}{\mcitedefaultseppunct}\relax
\EndOfBibitem
\bibitem[Ewald(1921)]{ewald1921berechnung}
Ewald,~P.~P. Die {Berechnung} optischer und elektrostatischer {Gitterpotentiale}. \emph{Annalen der Physik} \textbf{1921}, \emph{369}, 253--287\relax
\mciteBstWouldAddEndPuncttrue
\mciteSetBstMidEndSepPunct{\mcitedefaultmidpunct}
{\mcitedefaultendpunct}{\mcitedefaultseppunct}\relax
\EndOfBibitem
\bibitem[Zhou and Dai(2018)Zhou, and Dai]{zhou2018canonical}
Zhou,~C.-C.; Dai,~W.-S. Canonical partition functions: ideal quantum gases, interacting classical gases, and interacting quantum gases. \emph{Journal of Statistical Mechanics: Theory and Experiment} \textbf{2018}, \emph{2018}, 023105\relax
\mciteBstWouldAddEndPuncttrue
\mciteSetBstMidEndSepPunct{\mcitedefaultmidpunct}
{\mcitedefaultendpunct}{\mcitedefaultseppunct}\relax
\EndOfBibitem
\bibitem[Barghathi \latin{et~al.}(2020)Barghathi, Yu, and Del~Maestro]{barghathi2020theory}
Barghathi,~H.; Yu,~J.; Del~Maestro,~A. Theory of noninteracting fermions and bosons in the canonical ensemble. \emph{Physical Review Research} \textbf{2020}, \emph{2}, 043206\relax
\mciteBstWouldAddEndPuncttrue
\mciteSetBstMidEndSepPunct{\mcitedefaultmidpunct}
{\mcitedefaultendpunct}{\mcitedefaultseppunct}\relax
\EndOfBibitem
\bibitem[Dornheim \latin{et~al.}(2025)Dornheim, Tolias, Moldabekov, and Vorberger]{dornheim2025eta}
Dornheim,~T.; Tolias,~P.; Moldabekov,~Z.~A.; Vorberger,~J. $\eta$-ensemble path integral {Monte Carlo} approach to the free energy of the warm dense electron gas and the uniform electron liquid. \emph{Physical Review Research} \textbf{2025}, \emph{7}, 023250\relax
\mciteBstWouldAddEndPuncttrue
\mciteSetBstMidEndSepPunct{\mcitedefaultmidpunct}
{\mcitedefaultendpunct}{\mcitedefaultseppunct}\relax
\EndOfBibitem
\bibitem[Caillol and Gilles(2000)Caillol, and Gilles]{caillol2000monte}
Caillol,~J.~M.; Gilles,~D. Monte {Carlo} simulations of the {Yukawa} one-component plasma. \emph{Journal of Statistical Physics} \textbf{2000}, \emph{100}, 933--947\relax
\mciteBstWouldAddEndPuncttrue
\mciteSetBstMidEndSepPunct{\mcitedefaultmidpunct}
{\mcitedefaultendpunct}{\mcitedefaultseppunct}\relax
\EndOfBibitem
\bibitem[Plummer \latin{et~al.}(2025)Plummer, Svensson, Gericke, Hollebon, Vinko, and Gregori]{plummer2025ionization}
Plummer,~D.; Svensson,~P.; Gericke,~D.~O.; Hollebon,~P.; Vinko,~S.~M.; Gregori,~G. Ionization calculations using classical molecular dynamics. \emph{Physical Review E} \textbf{2025}, \emph{111}, 015204\relax
\mciteBstWouldAddEndPuncttrue
\mciteSetBstMidEndSepPunct{\mcitedefaultmidpunct}
{\mcitedefaultendpunct}{\mcitedefaultseppunct}\relax
\EndOfBibitem
\bibitem[Fukuda and Nakamura(2012)Fukuda, and Nakamura]{fukuda2012non}
Fukuda,~I.; Nakamura,~H. Non-Ewald methods: theory and applications to molecular systems. \emph{Biophysical Reviews} \textbf{2012}, \emph{4}, 161--170\relax
\mciteBstWouldAddEndPuncttrue
\mciteSetBstMidEndSepPunct{\mcitedefaultmidpunct}
{\mcitedefaultendpunct}{\mcitedefaultseppunct}\relax
\EndOfBibitem
\bibitem[Yakub and Ronchi(2003)Yakub, and Ronchi]{yakub2003efficient}
Yakub,~E.; Ronchi,~C. An efficient method for computation of long-ranged {Coulomb} forces in computer simulation of ionic fluids. \emph{The Journal of Chemical Physics} \textbf{2003}, \emph{119}, 11556--11560\relax
\mciteBstWouldAddEndPuncttrue
\mciteSetBstMidEndSepPunct{\mcitedefaultmidpunct}
{\mcitedefaultendpunct}{\mcitedefaultseppunct}\relax
\EndOfBibitem
\bibitem[Yakub and Ronchi(2005)Yakub, and Ronchi]{yakub2005new}
Yakub,~E.; Ronchi,~C. A new method for computation of long ranged {Coulomb} forces in computer simulation of disordered systems. \emph{Journal of Low Temperature Physics} \textbf{2005}, \emph{139}, 633--643\relax
\mciteBstWouldAddEndPuncttrue
\mciteSetBstMidEndSepPunct{\mcitedefaultmidpunct}
{\mcitedefaultendpunct}{\mcitedefaultseppunct}\relax
\EndOfBibitem
\bibitem[Yakub \latin{et~al.}(2007)Yakub, Ronchi, and Staicu]{yakub2007molecular}
Yakub,~E.; Ronchi,~C.; Staicu,~D. Molecular dynamics simulation of premelting and melting phase transitions in stoichiometric uranium dioxide. \emph{The Journal of Chemical Physics} \textbf{2007}, \emph{127}, 094508\relax
\mciteBstWouldAddEndPuncttrue
\mciteSetBstMidEndSepPunct{\mcitedefaultmidpunct}
{\mcitedefaultendpunct}{\mcitedefaultseppunct}\relax
\EndOfBibitem
\bibitem[Demyanov and Levashov(2022)Demyanov, and Levashov]{demyanov2022one}
Demyanov,~G.~S.; Levashov,~P.~R. One-component plasma of a million particles via angular-averaged {Ewald} potential: {A Monte Carlo} study. \emph{Physical Review E} \textbf{2022}, \emph{106}, 015204\relax
\mciteBstWouldAddEndPuncttrue
\mciteSetBstMidEndSepPunct{\mcitedefaultmidpunct}
{\mcitedefaultendpunct}{\mcitedefaultseppunct}\relax
\EndOfBibitem
\bibitem[Demyanov \latin{et~al.}(2024)Demyanov, Onegin, and Levashov]{demyanov2024N}
Demyanov,~G.; Onegin,~A.; Levashov,~P. $N$-convergence in one--component plasma: {Comparison} of {Coulomb}, {Ewald}, and angular--averaged {Ewald} potentials. \emph{Contributions to Plasma Physics} \textbf{2024}, \emph{64}, e202300164\relax
\mciteBstWouldAddEndPuncttrue
\mciteSetBstMidEndSepPunct{\mcitedefaultmidpunct}
{\mcitedefaultendpunct}{\mcitedefaultseppunct}\relax
\EndOfBibitem
\bibitem[Filinov \latin{et~al.}(2020)Filinov, Larkin, and Levashov]{filinov2020uniform}
Filinov,~V.~S.; Larkin,~A.~S.; Levashov,~P.~R. Uniform electron gas at finite temperature by fermionic-path-integral {Monte Carlo} simulations. \emph{Physical Review E} \textbf{2020}, \emph{102}, 033203\relax
\mciteBstWouldAddEndPuncttrue
\mciteSetBstMidEndSepPunct{\mcitedefaultmidpunct}
{\mcitedefaultendpunct}{\mcitedefaultseppunct}\relax
\EndOfBibitem
\bibitem[Dornheim \latin{et~al.}(2025)Dornheim, Chuna, Bellenbaum, Moldabekov, Tolias, and Vorberger]{dornheim2025application}
Dornheim,~T.; Chuna,~T.~M.; Bellenbaum,~H.~M.; Moldabekov,~Z.; Tolias,~P.; Vorberger,~J. Application of a spherically averaged pair potential in \textit{ab initio} path integral {Monte Carlo} simulations of the warm dense electron gas. \emph{arXiv preprint arXiv:2504.00737} \textbf{2025}, \relax
\mciteBstWouldAddEndPunctfalse
\mciteSetBstMidEndSepPunct{\mcitedefaultmidpunct}
{}{\mcitedefaultseppunct}\relax
\EndOfBibitem
\bibitem[Demyanov and Levashov(2022)Demyanov, and Levashov]{demyanov2022systematic}
Demyanov,~G.~S.; Levashov,~P.~R. Systematic derivation of angular-averaged {Ewald} potential. \emph{Journal of Physics A: Mathematical and Theoretical} \textbf{2022}, \emph{55}, 385202\relax
\mciteBstWouldAddEndPuncttrue
\mciteSetBstMidEndSepPunct{\mcitedefaultmidpunct}
{\mcitedefaultendpunct}{\mcitedefaultseppunct}\relax
\EndOfBibitem
\bibitem[Dornheim \latin{et~al.}(2024)Dornheim, Böhme, and Schwalbe]{dornheim2024ISHTAR}
Dornheim,~T.; Böhme,~M.; Schwalbe,~S. {ISHTAR} - {Imaginary-time Stochastic High-performance Tool for Ab initio Research}. 2024; \url{https://doi.org/10.5281/zenodo.10497098}\relax
\mciteBstWouldAddEndPuncttrue
\mciteSetBstMidEndSepPunct{\mcitedefaultmidpunct}
{\mcitedefaultendpunct}{\mcitedefaultseppunct}\relax
\EndOfBibitem
\bibitem[Mezzacapo and Boninsegni(2007)Mezzacapo, and Boninsegni]{mezzacapo2007structure}
Mezzacapo,~F.; Boninsegni,~M. Structure, superfluidity, and quantum melting of hydrogen clusters. \emph{Physical Review A} \textbf{2007}, \emph{75}, 033201\relax
\mciteBstWouldAddEndPuncttrue
\mciteSetBstMidEndSepPunct{\mcitedefaultmidpunct}
{\mcitedefaultendpunct}{\mcitedefaultseppunct}\relax
\EndOfBibitem
\bibitem[Dornheim \latin{et~al.}(2021)Dornheim, B{\"o}hme, Militzer, and Vorberger]{dornheim2021ab}
Dornheim,~T.; B{\"o}hme,~M.; Militzer,~B.; Vorberger,~J. \textit{Ab initio} path integral {Monte Carlo} approach to the momentum distribution of the uniform electron gas at finite temperature without fixed nodes. \emph{Physical Review B} \textbf{2021}, \emph{103}, 205142\relax
\mciteBstWouldAddEndPuncttrue
\mciteSetBstMidEndSepPunct{\mcitedefaultmidpunct}
{\mcitedefaultendpunct}{\mcitedefaultseppunct}\relax
\EndOfBibitem
\bibitem[Boninsegni \latin{et~al.}(2006)Boninsegni, Prokof’ev, and Svistunov]{boninsegni2006worm_a}
Boninsegni,~M.; Prokof’ev,~N.; Svistunov,~B. Worm algorithm for continuous-space path integral {Monte Carlo} simulations. \emph{Physical Review Letters} \textbf{2006}, \emph{96}, 070601\relax
\mciteBstWouldAddEndPuncttrue
\mciteSetBstMidEndSepPunct{\mcitedefaultmidpunct}
{\mcitedefaultendpunct}{\mcitedefaultseppunct}\relax
\EndOfBibitem
\bibitem[Boninsegni \latin{et~al.}(2006)Boninsegni, Prokof'ev, and Svistunov]{boninsegni2006worm_b}
Boninsegni,~M.; Prokof'ev,~N.~V.; Svistunov,~B.~V. Worm algorithm and diagrammatic {Monte Carlo}: {A} new approach to continuous-space path integral {Monte Carlo} simulations. \emph{Physical Review E} \textbf{2006}, \emph{74}, 036701\relax
\mciteBstWouldAddEndPuncttrue
\mciteSetBstMidEndSepPunct{\mcitedefaultmidpunct}
{\mcitedefaultendpunct}{\mcitedefaultseppunct}\relax
\EndOfBibitem
\bibitem[Dornheim \latin{et~al.}(2019)Dornheim, Groth, Filinov, and Bonitz]{dornheim2019path}
Dornheim,~T.; Groth,~S.; Filinov,~A.~V.; Bonitz,~M. Path integral {Monte Carlo} simulation of degenerate electrons: {Permutation-cycle} properties. \emph{The Journal of Chemical Physics} \textbf{2019}, \emph{151}, 014108\relax
\mciteBstWouldAddEndPuncttrue
\mciteSetBstMidEndSepPunct{\mcitedefaultmidpunct}
{\mcitedefaultendpunct}{\mcitedefaultseppunct}\relax
\EndOfBibitem
\bibitem[Groth \latin{et~al.}(2017)Groth, Dornheim, Sjostrom, Malone, Foulkes, and Bonitz]{groth2017ab}
Groth,~S.; Dornheim,~T.; Sjostrom,~T.; Malone,~F.~D.; Foulkes,~W. M.~C.; Bonitz,~M. \textit{Ab initio} exchange-correlation free energy of the uniform electron gas at warm dense matter conditions. \emph{Physical Review Letters} \textbf{2017}, \emph{119}, 135001\relax
\mciteBstWouldAddEndPuncttrue
\mciteSetBstMidEndSepPunct{\mcitedefaultmidpunct}
{\mcitedefaultendpunct}{\mcitedefaultseppunct}\relax
\EndOfBibitem
\bibitem[Thorpe \latin{et~al.}(2025)Thorpe, Smith, Hasnip, and Drummond]{thorpe2025acceleration}
Thorpe,~B.; Smith,~M.~J.; Hasnip,~P.~J.; Drummond,~N.~D. Acceleration of the {CASINO} quantum {Monte Carlo} software using graphics processing units and {OpenACC}. 2025; \url{https://arxiv.org/abs/2507.02888}\relax
\mciteBstWouldAddEndPuncttrue
\mciteSetBstMidEndSepPunct{\mcitedefaultmidpunct}
{\mcitedefaultendpunct}{\mcitedefaultseppunct}\relax
\EndOfBibitem
\bibitem[M{\"u}ller \latin{et~al.}(2023)M{\"u}ller, Christiansen, Schnabel, and Janke]{muller2023fast}
M{\"u}ller,~F.; Christiansen,~H.; Schnabel,~S.; Janke,~W. Fast, hierarchical, and adaptive algorithm for {Metropolis Monte Carlo} simulations of long-range interacting systems. \emph{Physical Review X} \textbf{2023}, \emph{13}, 031006\relax
\mciteBstWouldAddEndPuncttrue
\mciteSetBstMidEndSepPunct{\mcitedefaultmidpunct}
{\mcitedefaultendpunct}{\mcitedefaultseppunct}\relax
\EndOfBibitem
\bibitem[John \latin{et~al.}(2016)John, Spura, Habershon, and K{\"u}hne]{john2016quantum}
John,~C.; Spura,~T.; Habershon,~S.; K{\"u}hne,~T.~D. Quantum ring-polymer contraction method: {Including} nuclear quantum effects at no additional computational cost in comparison to \textit{ab initio} molecular dynamics. \emph{Physical Review E} \textbf{2016}, \emph{93}, 043305\relax
\mciteBstWouldAddEndPuncttrue
\mciteSetBstMidEndSepPunct{\mcitedefaultmidpunct}
{\mcitedefaultendpunct}{\mcitedefaultseppunct}\relax
\EndOfBibitem
\bibitem[Ceperley and Alder(1980)Ceperley, and Alder]{ceperley1980ground}
Ceperley,~D.~M.; Alder,~B.~J. Ground state of the electron gas by a stochastic method. \emph{Physical Review Letters} \textbf{1980}, \emph{45}, 566\relax
\mciteBstWouldAddEndPuncttrue
\mciteSetBstMidEndSepPunct{\mcitedefaultmidpunct}
{\mcitedefaultendpunct}{\mcitedefaultseppunct}\relax
\EndOfBibitem
\bibitem[Loos and Gill(2016)Loos, and Gill]{loos2016uniform}
Loos,~P.-F.; Gill,~P. M.~W. The uniform electron gas. \emph{Wiley Interdisciplinary Reviews: Computational Molecular Science} \textbf{2016}, \emph{6}, 410--429\relax
\mciteBstWouldAddEndPuncttrue
\mciteSetBstMidEndSepPunct{\mcitedefaultmidpunct}
{\mcitedefaultendpunct}{\mcitedefaultseppunct}\relax
\EndOfBibitem
\bibitem[Giuliani and Vignale(2008)Giuliani, and Vignale]{giuliani2008quantum}
Giuliani,~G.; Vignale,~G. \emph{Quantum Theory of the Electron Liquid}; Cambridge University Press: Cambridge, 2008\relax
\mciteBstWouldAddEndPuncttrue
\mciteSetBstMidEndSepPunct{\mcitedefaultmidpunct}
{\mcitedefaultendpunct}{\mcitedefaultseppunct}\relax
\EndOfBibitem
\bibitem[Dornheim \latin{et~al.}(2023)Dornheim, Tolias, Moldabekov, and Vorberger]{dornheim2023energy}
Dornheim,~T.; Tolias,~P.; Moldabekov,~Z.~A.; Vorberger,~J. {Energy response and spatial alignment of the perturbed electron gas}. \emph{The Journal of Chemical Physics} \textbf{2023}, \emph{158}, 164108\relax
\mciteBstWouldAddEndPuncttrue
\mciteSetBstMidEndSepPunct{\mcitedefaultmidpunct}
{\mcitedefaultendpunct}{\mcitedefaultseppunct}\relax
\EndOfBibitem
\bibitem[Kasim and Vinko(2021)Kasim, and Vinko]{PhysRevLett.127.126403}
Kasim,~M.~F.; Vinko,~S.~M. Learning the Exchange-Correlation Functional from Nature with Fully Differentiable Density Functional Theory. \emph{Phys. Rev. Lett.} \textbf{2021}, \emph{127}, 126403\relax
\mciteBstWouldAddEndPuncttrue
\mciteSetBstMidEndSepPunct{\mcitedefaultmidpunct}
{\mcitedefaultendpunct}{\mcitedefaultseppunct}\relax
\EndOfBibitem
\bibitem[Dornheim \latin{et~al.}(2024)Dornheim, Tolias, Moldabekov, and Vorberger]{dornheim2024eta}
Dornheim,~T.; Tolias,~P.; Moldabekov,~Z.; Vorberger,~J. {$\eta$}-ensemble path integral {Monte Carlo} approach to the free energy of the warm dense electron gas and the uniform electron liquid. \emph{arXiv preprint arXiv:2412.13596} \textbf{2024}, \relax
\mciteBstWouldAddEndPunctfalse
\mciteSetBstMidEndSepPunct{\mcitedefaultmidpunct}
{}{\mcitedefaultseppunct}\relax
\EndOfBibitem
\bibitem[Metropolis \latin{et~al.}(1953)Metropolis, Rosenbluth, Rosenbluth, Teller, and Teller]{metropolis1953equation}
Metropolis,~N.; Rosenbluth,~A.~W.; Rosenbluth,~M.~N.; Teller,~A.~H.; Teller,~E. Equation of state calculations by fast computing machines. \emph{The Journal of Chemical Physics} \textbf{1953}, \emph{21}, 1087--1092\relax
\mciteBstWouldAddEndPuncttrue
\mciteSetBstMidEndSepPunct{\mcitedefaultmidpunct}
{\mcitedefaultendpunct}{\mcitedefaultseppunct}\relax
\EndOfBibitem
\bibitem[Hastings(1970)]{hastings1970monte}
Hastings,~W.~K. Monte {Carlo} sampling methods using {Markov} chains and their applications. \emph{Biometrika} \textbf{1970}, \emph{57}, 97--109\relax
\mciteBstWouldAddEndPuncttrue
\mciteSetBstMidEndSepPunct{\mcitedefaultmidpunct}
{\mcitedefaultendpunct}{\mcitedefaultseppunct}\relax
\EndOfBibitem
\bibitem[Dornheim \latin{et~al.}(2016)Dornheim, Groth, Sjostrom, Malone, Foulkes, and Bonitz]{dornheim2016ab}
Dornheim,~T.; Groth,~S.; Sjostrom,~T.; Malone,~F.~D.; Foulkes,~W. M.~C.; Bonitz,~M. \textit{Ab initio} quantum {Monte Carlo} simulation of the warm dense electron gas in the thermodynamic limit. \emph{Physical Review Letters} \textbf{2016}, \emph{117}, 156403\relax
\mciteBstWouldAddEndPuncttrue
\mciteSetBstMidEndSepPunct{\mcitedefaultmidpunct}
{\mcitedefaultendpunct}{\mcitedefaultseppunct}\relax
\EndOfBibitem
\bibitem[Not()]{Note-1}
See \url{https://github.com/fdmalone/uegpy}\relax
\mciteBstWouldAddEndPuncttrue
\mciteSetBstMidEndSepPunct{\mcitedefaultmidpunct}
{\mcitedefaultendpunct}{\mcitedefaultseppunct}\relax
\EndOfBibitem
\bibitem[Singwi \latin{et~al.}(1968)Singwi, Tosi, Land, and Sj{\"o}lander]{singwi1968electron}
Singwi,~K.; Tosi,~M.; Land,~R.; Sj{\"o}lander,~A. Electron correlations at metallic densities. \emph{Physical Review} \textbf{1968}, \emph{176}, 589\relax
\mciteBstWouldAddEndPuncttrue
\mciteSetBstMidEndSepPunct{\mcitedefaultmidpunct}
{\mcitedefaultendpunct}{\mcitedefaultseppunct}\relax
\EndOfBibitem
\bibitem[Caillol and Gilles(2010)Caillol, and Gilles]{caillol2010accurate}
Caillol,~J.-M.; Gilles,~D. An accurate equation of state for the one-component plasma in the low coupling regime. \emph{Journal of Physics A: Mathematical and Theoretical} \textbf{2010}, \emph{43}, 105501\relax
\mciteBstWouldAddEndPuncttrue
\mciteSetBstMidEndSepPunct{\mcitedefaultmidpunct}
{\mcitedefaultendpunct}{\mcitedefaultseppunct}\relax
\EndOfBibitem
\bibitem[Sakkos \latin{et~al.}(2009)Sakkos, Casulleras, and Boronat]{sakkos_JCP_2009}
Sakkos,~K.; Casulleras,~J.; Boronat,~J. High order Chin actions in path integral Monte Carlo. \emph{The Journal of Chemical Physics} \textbf{2009}, \emph{130}, 204109\relax
\mciteBstWouldAddEndPuncttrue
\mciteSetBstMidEndSepPunct{\mcitedefaultmidpunct}
{\mcitedefaultendpunct}{\mcitedefaultseppunct}\relax
\EndOfBibitem
\bibitem[Dornheim \latin{et~al.}(2015)Dornheim, Groth, Filinov, and Bonitz]{Dornheim_NJP_2015}
Dornheim,~T.; Groth,~S.; Filinov,~A.; Bonitz,~M. Permutation blocking path integral Monte Carlo: a highly efficient approach to the simulation of strongly degenerate non-ideal fermions. \emph{New Journal of Physics} \textbf{2015}, \emph{17}, 073017\relax
\mciteBstWouldAddEndPuncttrue
\mciteSetBstMidEndSepPunct{\mcitedefaultmidpunct}
{\mcitedefaultendpunct}{\mcitedefaultseppunct}\relax
\EndOfBibitem
\end{mcitethebibliography}

\end{document}